\documentclass[review,3p]{elsarticle}

\usepackage{amssymb}
\usepackage{amsmath}
\journal{Journal of Nonlinear Science}
\newcommand{\virgolette}[1]{``#1''}
\usepackage{epsfig}
\usepackage{graphics}
\usepackage{chemfig}
\usepackage{amsmath,bm}
\usepackage{hyperref}
\usepackage{tabularx}
\usepackage{makecell}
\usepackage{booktabs}
\usepackage{multirow}

\usepackage{float}

\hypersetup{
 colorlinks = true,
 citecolor= blue,
 linkcolor= blue,
 urlcolor= blue
}
\usepackage{comment}
\usepackage{color}
\usepackage{xcolor}
\DeclareMathOperator*{\argmax}{arg\,max}
\DeclareMathOperator*{\argmin}{arg\,min}
\newcommand{\LB}[1]{{\color{blue} #1}}
\newcommand{\corr}{\color{purple} }
\newcommand{\corrB}{\color{blue} }
\newcommand{\corrPB}{\color{red} }
\newcommand{\ptpt}[1]{\left( #1 \right)}
\newcommand{\pqpq}[1]{\left[ #1 \right]}
\newcommand{\pgpg}[1]{\left\lbrace #1 \right\rbrace}
\newcommand{\frapar}[2]{\dfrac{\partial #1}{\partial #2}}
\newcommand{\fraparsq}[2]{\dfrac{\partial ^2 #1}{\partial #2^2}}
\newcommand{\fraparcu}[2]{\dfrac{\partial ^3 #1}{\partial #2^3}}
\usepackage[normalem]{ulem}

\begin{document}

\begin{frontmatter}

%\author[1]{Bo Lin}\ead{matboln@nus.edu.sg}
\author[1]{Pierpaolo Belardinelli\corref{cor1}} \ead{p.belardinelli@univpm.it}
\cortext[cor1]{Corresponding author}

%\affiliation[1]{organization={Department of Mathematics, National University of Singapore},
% addressline={Lower Kent Ridge Road},
 % city={Singapore}, 
% postcode={119076}, country={Singapore}}
\affiliation[1]{organization={DICEA, Polytechnic University of Marche}, city={Ancona},
          postcode={60131},
          country={Italy}}

\title{Quasiperiodicity breakdown and homoclinic bifurcations in the forced librations of a parametric oscillator
%Quasiperiodicity breakdown and homoclinic bifurcations at the libration resonances of a parametric oscillator
%Quasiperiodicity breakdown and homoclinic bifurcations under fast-slow parametric excitation  
%in a parametrically-driven Duffing oscillator under slow parametric modulation 
}

%%%% Abstract text to be placed here %%%%%%%%%%%%
\begin{abstract} 
Above its instability threshold, a parametrically excited oscillator settles into one of two phase states. The response of these states to a second, much slower modulation of the same stiffness is investigated. In the rotating frame, this modulation periodically varies the detuning and drives the phase librations. When the modulation is slow compared with the libration frequency, the phase states follow adiabatically. As the two frequencies become comparable, this approximation breaks down and the driven libration resonates with the intrinsic one. This nonadiabatic regime is examined for a Duffing oscillator with  nonlinear damping. The averaged equations describe a periodically perturbed planar system with integrable unperturbed dynamics, whose homoclinic loop is derived in closed form. The phase states persist as invariant tori supporting quasiperiodic motion. Their stability and period-doubling bifurcations are captured by a damped Mathieu equation, in which the modulation contributes to both parametric stiffness and parametric damping. A Melnikov analysis yields explicit thresholds for separatrix splitting and the onset of homoclinic intersections, which are corroborated by numerical computations of the invariant manifolds. Numerically it is revealed that these intersections arise well before sustained chaotic motion develops. The fate of the parametric oscillation depends markedly on the resonance condition: modulation at the libration frequency destroys the invariant tori and quenches the oscillation, whereas modulation at twice the libration frequency leaves it intact.

%Unconventional multifrequency driving gives rise to rich dynamics beyond the scope of single-tone excitation. This study examines the nonlinear response of a Duffing resonator under a combined fast-and-slow parametric drive. By applying the Melnikov technique, we analytically characterize the transverse homoclinic orbits and establish the threshold for chaotic motion.
\end{abstract}  
%%%%%%%%%%%%%%%%%%%%%%%%%%%

\end{frontmatter}

\section{Introduction}
Linear vibration theory holds few secrets: superposition applies, and
every response can be traced back to its cause. Nonlinear behavior, on
the other hand, continues to surprise with unexpected and interesting
dynamics. Examples are numerous   in the realm of micro- and nanomechanics, where nonlinearities
are not the exception but the rule
\cite{LifshitzCross2008,Bachtold2022,Steeneken2021,Belardinelli2025}.
 
An emerging direction in this field moves beyond the single-harmonic excitation on which much of our intuition is built, exploring instead the effects of multiple driving tones. Modulating the amplitude or frequency of the excitation has, for instance, been exploited to generate chaos in micromechanical resonators, with applications to chaotic signal generation and random-number generation \cite{Defoort2021}. Even two detuned forces acting on a Duffing resonator can give rise to dynamics that defy single-tone theory, including transitions between coexisting low- and high-amplitude states \cite{Catalini2025}.  In a frame rotating with the primary tone, the secondary tone appears as a modulation of the phase-space flow. 
The stable responses to the primary excitation correspond to fixed points in this frame, small perturbations undergo oscillatory motion known as librations, a term borrowed from celestial mechanics \cite{Howell1997}.
The libration frequency determines the satellite peaks observed around the carrier in the spectrum of a driven nonlinear mode \cite{Huber2020,Ochs2021}. Feedback can even sustain these oscillations, turning the resonator into a librator \cite{Houri2021}. 
The libration frequency therefore provides an intrinsic dynamical timescale, distinct from the relaxation time governing the approach to the stationary states.

When the beat period is long compared with the ring-down time, the state
follows the drifting attractor adiabatically; as the beat frequency
increases, it lags behind \cite{Catalini2025}.  The onset of
inter-attractor transitions has been characterised as a dynamical phase
transition, whose location is set not by the ring-down time alone but by
the resonance properties of the stationary states \cite{Kumar2026}.  Two-tone excitation has also been proposed for sensor readout \cite{Houri2019}, while a separate rotating-frame approach has been developed as a generic route to chaos in nonlinear micro- and nanoelectromechanical resonators \cite{Houri2020}.

In these studies, the additional tones enter through the applied force. A different situation arises when a secondary tone modulates an intrinsic parameter of the resonator, such as its stiffness, rather than adding an external force. An example is the modulation of the tension, and hence of the stiffness, of a resonator subject to a primary resonant drive \cite{wattjes2026}. 
Related nonlinear analyses of combined parametric and external excitation have also been developed; for example, Yagasaki et al. used averaging and Melnikov methods to establish invariant tori and chaotic dynamics in a weakly nonlinear single-degree-of-freedom system \cite{Yagasaki1990}.
An especially interesting case occurs when the primary excitation is itself parametric.

Above the parametric instability threshold, a parametrically pumped oscillator possesses two stable phase states, related by a phase difference of $\pi$. Together with the threshold separating the subcritical and postcritical regimes, these states have made the parametric oscillator a versatile platform. 
%a versatile platform. These two ingredients, a threshold and a pair of degenerate phase states, have made the parametric oscillator a working tool.  
Below threshold, it amplifies one quadrature and squeezes noise in the other \cite{Rugar1991}. Above threshold, the phase states serve as bits \cite{Mahboob2008} and can be used to emulate Ising Hamiltonians \cite{Mahboob2016}, Boltzmann machines \cite{Margiani2025}, and qubits \cite{Grimm2020}. Beyond these stationary states, recent experiments have also demonstrated noise-activated switching between coexisting limit-cycle attractors emerging from parametrically driven resonators \cite{margiani2026activatedswitchingcoexistinglimit}. Nonlinear damping gives parametric oscillators better short-term frequency stability than their directly driven counterparts \cite{Kartal2026}.

Consider now a parametrically pumped oscillator whose stiffness is subject to a secondary, much slower modulation. In the frame rotating at half the pump frequency, this modulation periodically varies the effective detuning and directly drives the phase dynamics. 
%The secondary modulation   therefore interacts with these intrinsic oscillations.  
%In the absence of the secondary modulation, the two phase states are fixed points of the averaged dynamics, and perturbations around either state undergo librations. The secondary modulation can therefore interact with these intrinsic oscillations. When it is sufficiently slow compared with the characteristic timescales of the rotating-frame dynamics, the phase states can follow the slowly varying system adiabatically. As the modulation frequency approaches the natural libration frequency, however, a resonance develops between the imposed modulation and the intrinsic phase oscillations, and the adiabatic description breaks down. THIS IS COOL BUT MAY BE A REPETITION
Earlier work on two-tone parametric excitation falls into two categories. The first  is mostly linear: two stiffness modulations of incommensurate frequency lead to the quasiperiodic Mathieu equation, whose stability charts are far more intricate than the classical tongues
\cite{ZounesRand1998,Rand2003,Kovacic2018}; a nonlinear version has
been treated via Lie transform perturbation theory with elliptic functions \cite{Zounes2002}. The second
is nonlinear and adiabatic. For a shallow arch, delayed bifurcations and the sequence of equilibria visited during a slow parametric excitation cycle have been analysed, with chaotic regions also identified through Melnikov analysis \cite{LAKRAD20041149}. Related two-tone studies combined a resonant or fast excitation with a much slower parametric modulation and obtained conditions for quasiperiodic solutions and periodic bursters \cite{LAKRAD2005813,LAKRAD20092426}. Under resonant external forcing and slow parametric support motion, bursters involve fixed-point, quasiperiodic and chaotic states, and the slow parametric excitation was shown to suppress chaos induced by the resonant forcing \cite{Lakrad2016}. Under fast–slow parametric excitation, quasiperiodic bursters were reported, including period-doubling and period-tripling sequences that lead to hysteretic chaos. The fast excitation can also suppress chaos induced by the slow modulation \cite{Chtouki2020}. 
%These studies primarily address the slow-modulation regime, in which the response can be described in terms of slow evolution along branches of the averaged dynamics and delayed transitions at bifurcations.
These studies primarily address dynamics arising from a pronounced separation of timescales, in which the slow parametric excitation can be treated adiabatically or within a slow–fast framework. A complementary regime emerges when the modulation, although still slow compared with the carrier oscillation, becomes comparable to the libration/relaxation timescales of the dynamics in the rotating frame.  
This nonadiabatic regime has recently been explored experimentally in a graphene resonator subject to a resonant direct drive and a slow modulation of its membrane tension \cite{wattjes2026}. When the modulation rate becomes comparable to the mechanical relaxation rate, adiabatic tracking of the stationary states breaks down. Parametric instabilities then emerge in the slow dynamics at frequencies determined by the libration around these states, rather than by the mechanical resonance. Stronger modulation produces period doubling and chaos. 
%A slow-variable model and Melnikov analysis were used to describe the instability mechanisms and the onset of homoclinic intersections.

The present investigation considers a different configuration, in which the primary excitation is itself parametric rather than a resonant direct drive. 
The question addressed here is how secondary stiffness modulation excites librations around the parametric phase states, affects the stability of the resulting invariant tori, and modifies the global phase-space structure, in particular the homoclinic separatrix. 
The averaged equations retain the zero-amplitude equilibrium, which loses stability at the parametric threshold and gives rise to two symmetry-related, finite-amplitude phase states. Moreover, both linear and nonlinear damping contribute to the post-critical dynamics. These features fundamentally change the phase-space structure compared with that of a directly driven Duffing resonator. 

The analysis yields three main results. First, the forced libration around a phase state is an invariant torus of the full system, on which the motion is quasiperiodic, and the saddle persists as a torus of saddle type. Both are approximated explicitly in this work. In this approximation the forced libration grows without bound as the modulation frequency approaches the libration frequency. This primary resonance marks the end of the adiabatic picture.

The torus loses stability by period doubling. Its variational equation is a damped Mathieu equation in which the modulation acts at once as parametric stiffness and as parametric damping, a combination whose relative phase is known to decide between amplification and suppression \cite{Ardister2024}. The instability tongue is centred at twice the libration frequency. A Liouville transformation shows that the damping modulation feeds a second harmonic into the effective stiffness, and numerical continuation reveals a boundary markedly richer than the classical Mathieu tongue. The rest state has a tongue of its own, obtained here from Floquet theory.

Finally, analytical thresholds are obtained as functions of the modulation frequency for the transverse intersection of the stable and unstable manifolds along the outer and inner branches of the separatrix, by evaluating the Melnikov integral   on the closed-form homoclinic orbits \cite{Melnikov1963,Guckenheimer2013}.  Numerical computation of the manifolds confirms these thresholds. The tangency creates a horseshoe well before chaos is observed. Stable periodic orbits continue to attract almost all initial conditions, and sustained chaos sets in only at considerably larger modulation depth, after the tori have broken down through the bifurcation sequences documented here for two modulation frequencies, one near the libration frequency and one near twice its value.

\section{Equation of motion and averaging}
\label{sec:model}
To gain physical insight into how a secondary stiffness modulation acts on the phase states of a parametric oscillator and to identify the key governing parameters, a minimal single-mode model of the oscillator is introduced:
\begin{equation}
\ddot{x}  + \tilde{\delta}\dot{x}  + \tilde{\alpha} x^3 +\tilde{\delta}_\text{\tiny NL}  x^2 \dot{x}+ \left(\omega_{\tiny 0}^2   +\tilde{\gamma}\cos\!\big(2\omega_\text{\tiny F} t\big) + \tilde{\beta}\cos\phi\right)x=  0
\label{eq:eom}
 \end{equation}
where $x$ is the displacement,   the overdot denotes differentiation with respect to time, $\omega_{\tiny 0}$ is the linear resonant frequency, $\omega_\text{\tiny F}$  is the pump frequency (the parametric pump acts at $2\omega_\text{\tiny F}$), and $\phi(t)=\omega_\beta t$ is the phase of the secondary modulation with $\omega_\beta t\ll \omega_\text{\tiny F}$
The parameters $\tilde{\gamma}$ and $\tilde{\beta}$ are the pump and modulation amplitude, $\tilde{\delta}$ and $\tilde{\delta}_\text{\tiny NL}$ are the linear and nonlinear damping coefficient, and $\tilde{\alpha}$ is the Duffing coefficient.

%The parameters $\tilde{\delta}$, $\tilde{\delta}_\text{\tiny NL}$, $\omega_{\tiny 0}$, and $\tilde{\alpha}$ are the linear damping coefficient, nonlinear damping coefficient, natural frequency of the resonator,  and Duffing coefficient, respectively. The quantities $\tilde{\gamma}$ and $\tilde{\beta}$  correspond to the amplitude of the fast and slow parametric excitation that is modulated at $\phi(t)=\omega_\beta t\ll \omega_{\tiny 0} t$. 
%Finally,   $\omega_\text{\tiny F}\approx  \omega_{\tiny 0}$ is the frequency of the parametric drive.
%and $\theta_F$ are the angular frequency and phase shift of the fast drive.
%The dynamics arising from Eq.~\eqref{eq:eom} in the absence of the slow modulation ($\tilde{\beta}=0$) dates back centuries \cite{Faraday1831}. Above a certain   amplitude of the parametric excitation, the system becomes unstable and exhibits   parametric resonance \cite{}. 

\begin{figure*}[htb!]
\centering
\includegraphics[width=1\textwidth]{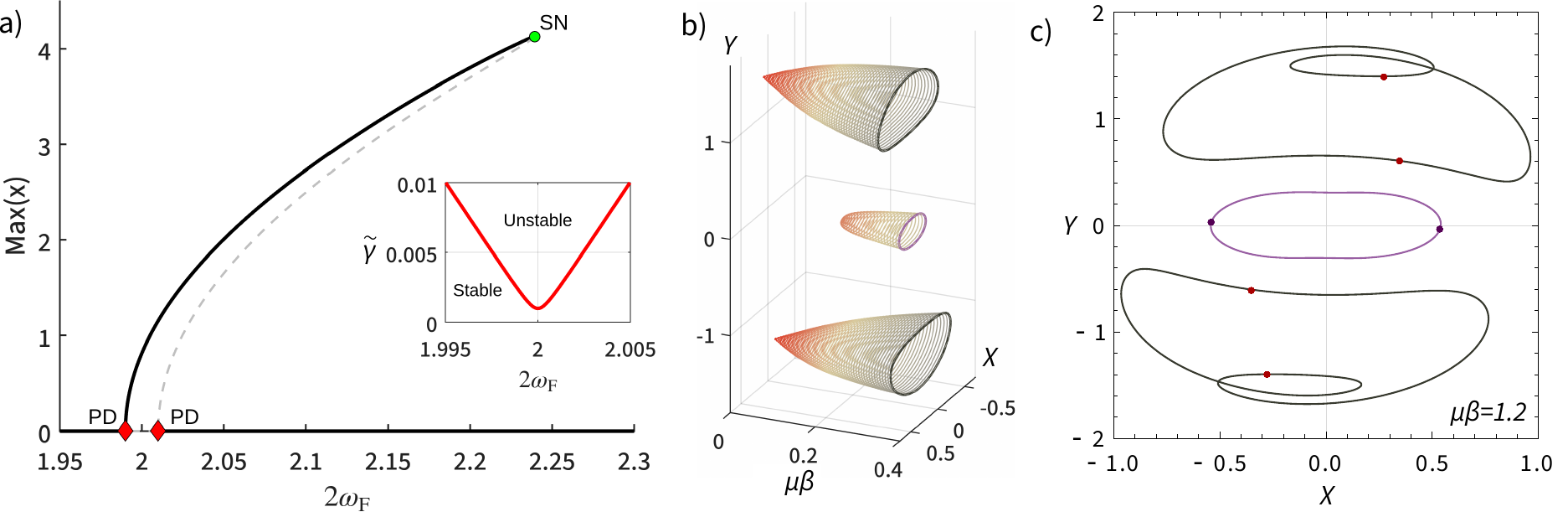}
\caption{
The parametric oscillator and its forced librations.
(a)~Frequency response without modulation ($\tilde\beta=0$): maximum
displacement versus pump frequency. Solid line: stable
phase-state branch; dashed line: unstable branch; the thick line on the
axis is the rest state. PD: period-doubling bifurcations of the rest
state (i.e. a pitchfork in the slow-flow); SN: saddle-node bifurcation. Inset: Ince-Strutt diagram, the
period-doubling locus plane, which
separates the stable from the unstable rest state. 
%The sketch shows the two excitations acting on the same parameter: the fast pump, of amplitude $\tilde\gamma$, and the slow modulation, of amplitude$\tilde\beta$. 
(b)~Invariant closed curves of the averaged system in
the $(X,Y)$ plane for increasing modulation depth $\mu\beta$ ($\nu_0=3.5$). The two curves originated from the phase-states exist
from $\mu\beta=0$; the curve around the rest state appears at finite
depth. Colour encodes $\mu\beta$. (c)~The three curves at $\mu\beta=1.2$. Dots are the stroboscopic points at the modulation period. Parameters: $\tilde{\delta}=0.00049505$, $\tilde{\alpha}=0.02$, $\tilde{\delta}_\text{\tiny NL}=0.0019802$, $\tilde{\beta}=0$.}
%Nonlinear dynamics of the parametrically-driven Duffing oscillator (Parameters are $\omega_{\tiny 0}=1$,  $\tilde{\delta}=0.00049505$, $\tilde{\alpha}=0.02$, $\tilde{\delta}_\text{\tiny NL}=0.0019802$, $\tilde{\beta}=0$). (a) Frequency response ($\tilde{\gamma}=0.02$).  PD and SN are Period-Doubling and Saddle-Node bifurcations. In the inset the  Ince-Strutt diagram, the  loci of the Period Doubling.  (b) Evolution of the orbit on the invariant tori in the $X-Y$ quadratures space. Color gradient is for increasing $\mu\beta\in[0,0.4]$. (c) In-phase and out-of-phase component $X$-$Y$ ($\mu\beta=1.2$, $\Omega=2.01$).}
\label{fig:overview}
\end{figure*}

In the absence of the secondary modulation $\tilde{\beta}=0$, Eq.~\eqref{eq:eom} describes a parametrically driven Duffing oscillator, a nonlinear extension of a classical model of parametric resonance. The phenomenon itself has a long history, dating back to Faraday's observation of subharmonic surface waves 
\cite{Faraday1831}. The mathematical framework for analysing such systems rests on the Mathieu-Hill equations \cite{Mathieu1868,Hill1886} and Floquet theory \cite{Floquet1883}. Above a certain amplitude of the parametric excitation, the zero-amplitude state becomes unstable and the system exhibits parametric resonance \cite{Kovacic2018}.  
The resulting dynamics for a fixed pump amplitude is shown in the bifurcation diagram of Fig.~\ref{fig:overview}(a). At the two frequencies marked by red diamonds, a period-doubling (PD)  bifurcation of the trivial state occurs:
%(i.e. a pitchfork  in the slow-flow): 
the zero-amplitude solution loses stability, and the system transitions to a finite-amplitude response (solid black curve). This gives rise to the two symmetry-related phase states, differing by a phase of \(\pi\), that are central to this work. As the frequency is further detuned, the stable branch terminates at a saddle-node (SN) bifurcation (green dot), beyond which the steady-state response folds back onto an unstable branch (dashed grey line). 

The parameter dependence of this instability is captured by the Arnold tongue shown in the inset of Fig.~\ref{fig:overview}(a). In the plane of the pump amplitude \(\tilde{\gamma}\) and the drive frequency \(2\omega_F\), the red boundary separates the stable and unstable regions: for parameters below the boundary, the zero-amplitude state is stable, whereas above it the system is unstable. The instability threshold is minimized at \(2\omega_F = 2\omega_0\), where the pump frequency matches the natural frequency of the oscillator.

 % Figure 1

The small parameters are introduced to distinguish the leading-order
dynamics from the weaker dissipative and modulation effects. In
particular, two perturbation parameters, $\epsilon$ and $\mu$, are
considered, with $0<\epsilon\ll\mu\ll 1$,   such that 
$\tilde{\gamma}=\epsilon\gamma$, 
$\tilde{\alpha}=\epsilon\alpha$, 
$\tilde{\beta}=\epsilon\mu\beta$, $\tilde{\delta}=\epsilon\mu\delta$, $\tilde{\delta}_\text{\tiny NL}=\epsilon\mu\delta_\text{\tiny NL}$.
Thus, the primary parametric excitation and cubic stiffness
nonlinearity enter at order $\epsilon$, whereas damping and the
secondary stiffness modulation are weaker, of order
$\epsilon\mu$.

%order the small terms using two small parameters $\epsilon$ and $\mu$ ruled as to $0<\epsilon\ll\mu\ll 1$, such that $\tilde{\gamma}=\epsilon\gamma$, $\tilde{\alpha}=\epsilon\alpha$,  $\tilde{\beta}=\epsilon\mu\beta$, $\tilde{\delta}=\epsilon\mu\delta$, $\tilde{\delta}_\text{\tiny NL}=\epsilon\mu\delta_\text{\tiny NL}$

Since the vibration frequency is much higher than all other rates and frequencies in the system, the vibrations can be described in a rotating frame. Introducing a slowly varying complex amplitude $z(\tau)$, the
displacement is expressed as
$x(t)=\dfrac{1}{2}\left(\overline{z}(\tau)e^{-i\omega_\text{F} t}+z(\tau)e^{i\omega_\text{F} t}\right)$,  with slow time variable $\tau=\epsilon t/2\omega_\text{\tiny F}$. 
 Substitution of this ansatz into Eq.~\eqref{eq:eom}, followed by
averaging over the fast oscillation period \cite{kryloff1947introduction}, eliminates the non-resonant terms and yields the slow-flow equation
\begin{equation}
    \dot{z} = \frac{3}{4} i  \alpha |z|^2z - \frac{1}{4}  \mu\delta_\text{\tiny NL} \overline{z}\omega_\text{\tiny F} z^2 + i \mu\beta \cos\phi_0 z - i \Omega z - \mu  \delta \omega_\text{\tiny F} z + \frac{1}{2} i \gamma \overline{z}
    \label{eq:complexeom}
\end{equation}
where overdot is differentiation with respect to $\tau$,  
and the scaled detuning and modulation phase are defined as
$\omega_\text{\tiny F}^2-\omega_{\tiny 0}^2=\epsilon \Omega$ and $\phi_0(\tau)=\nu_0 \tau$. Here, $\nu_0=2\omega_\text{\tiny F} \omega_\beta/\epsilon$ is the frequency of the secondary modulation on the slow timescale.  
Writing the complex amplitude in terms of its real and imaginary
parts, $z=X+iY$, with $X=\Re(z)$ and $Y=\Im(z)$, Eq.~\eqref{eq:complexeom}
can equivalently be expressed as a planar system of coupled
first-order differential equations:
%Since the vibration frequency is much higher than all other rates and frequencies in the system, the vibrations can be described in a rotating frame using a complex vibration amplitude $z$. To analyze the slow dynamics of Eq.~\eqref{eq:eom}, we employ the  averaging method \cite{kryloff1947introduction}. The asymptotic procedure substitutes a slowly varying complex-amplitude ansatz, i.e.,  $x(t)=\dfrac{1}{2}\left(\overline{z}(\tau)e^{-i\omega_\text{F} t}+z(\tau)e^{i\omega_\text{F} t}\right)$,  with slow time variable $\tau=\epsilon t/2\omega_\text{\tiny F}$, by averaging  over the oscillation period, thereby filtering out fast oscillatory terms, we obtain the complex equation 
%\begin{equation}
%    \dot{z} = \frac{3}{4} i \overline{z} \alpha z^2 - \frac{1}{4}  \mu\delta_\text{\tiny NL} \overline{z}\omega_\text{\tiny F} z^2 + i \mu\beta \cos\phi_0 z - i \Omega z - \mu  \delta \omega_\text{\tiny F} z + \frac{1}{2} i \gamma \overline{z}
%    \label{eq:complexeom}
%\end{equation}
%where overdot is differentiation with respect to $\tau$, and $\omega_\text{\tiny F}^2-\omega_{\tiny 0}^2=\epsilon \Omega$, $\phi_0(\tau)=\nu_0 \tau$,  $\nu_0=2\omega_\text{\tiny F} \omega_\beta/\epsilon$.
%Equation~\eqref{eq:complexeom} projected in  the quadrature space $(X=\Re(z),Y=\Im(z))$, reduces the original second-order dynamics to the coupled first-order differential equations: 
\begin{equation}
\left\{
\begin{aligned}
\dot{X} &= \Omega Y
          - \frac{3}{4} \alpha   (X^2+Y^2) \, Y
          - \mu\delta  \omega_\text{\tiny F} X  
          + \frac{\gamma}{2}  Y
          -\mu\frac{\delta_\text{\tiny NL} \omega_\text{\tiny F}}{4}    (X^2+Y^2)\, X- \mu\beta \, Y\cos\nu_0 \tau  , \\
\dot{Y} &= - \Omega X+\frac{3}{4} \alpha (X^2+Y^2)\, X
          - \mu\delta \omega_\text{\tiny F} Y
          + \frac{\gamma}{2}  X
          - \mu\frac{\delta_\text{\tiny NL} \omega_\text{\tiny F}}{4}   (X^2+Y^2)\, Y + \mu\beta \, X\cos\nu_0 \tau.
          \label{eq:XY}
\end{aligned}
\right.
\end{equation}

For $\beta=0$, the averaged system is autonomous and its stationary responses correspond to fixed points in the quadrature $(X,Y)$ plane.
For $\beta\neq0$, the secondary modulation makes the slow-flow equations explicitly time-periodic, replacing these fixed points with modulation-driven trajectories. Figure~\ref{fig:overview}(b)
illustrates the evolution of the corresponding orbits  as the modulation amplitude $\mu\beta$ increases. 
Besides the trajectories associated with the two finite-amplitude
phase states, a central orbit develops around the origin, indicating
the emergence of a stable finite-amplitude response associated with the otherwise unstable zero-amplitude state. In the original oscillator, all the modulated responses correspond to quasiperiodic
motion.
At a larger modulation amplitude, $\mu\beta=1.2$, the stroboscopic
Poincaré map shown in Fig.~\ref{fig:overview}(c) exhibits two distinct points for each response, indicating that the corresponding orbits have doubled their period. This observation motivates a detailed investigation of the stability of the modulation-driven responses and the bifurcations through which they lose stability. In particular, the corresponding invariant tori of the full system and their period-doubling bifurcations will be examined.

Via the transformation 
 \begin{equation}
 J=\dfrac{X^2+Y^2}{2},\quad \psi=\tan^{-1}(Y/X),
\end{equation}
 the  physical quadrature system~\eqref{eq:XY}  is written into the  canonical    action-angle   framework 
 \begin{equation}
\left\{
\begin{aligned}
\dot{J} &=  \gamma J \sin(2\psi)- 2 \mu\delta  \omega_\text{\tiny F} J -\mu\delta_\text{\tiny NL}  \omega_\text{\tiny F}  J^2  , \\
\dot{\psi} &=- \Omega + \frac{3 }{2} \alpha J + \frac{\gamma}{2}  \cos(2\psi) + \mu\beta \cos\nu_0 \tau.
\end{aligned}
\right.
\label{eq:JdotPhidot}
\end{equation}
that uncovers its Hamiltonian structure and explicitly expose the homoclinic separatrix loop.  
The unperturbed system of Eq.~\eqref{eq:JdotPhidot} with $\mu = 0$, 
is Hamiltonian with Hamiltonian energy 
\begin{equation}
H(J,\psi) = \frac{3}{4}\alpha J^2 - \Omega J + \frac{\gamma}{2}J\cos 2\psi.%, \qquad a = \dfrac{3}{4}\alpha. 
\label{eq:Hamiltonian}
\end{equation} 
Hamilton's equations give
\begin{equation}
\left\{
\begin{aligned}
\dot{J} &=-\frac{\partial H}{\partial \psi}=  \gamma J \sin(2\psi), \\
\dot{\psi} &= \frac{\partial H}{\partial J} =- \Omega + \frac{3}{2}\alpha  J + \frac{\gamma}{2}  \cos(2\psi).
\end{aligned}
\right.
\label{eq:systemHamiltonian}
\end{equation}
%\[
%\dot J = -\frac{\partial H}{\partial \phi} = F_1 J \sin 2\phi, \qquad
%\dot \phi = \frac{\partial H}{\partial J} = 2aJ - \Omega + \frac{F_1}{2}\cos 2\phi,
%\] which are the undamped versions of \eqref{eq:JdotPhidot}.
Assuming $\alpha>0$, the non trivial fixed points $p_i=(j_i,\psi_i)$ are centers,  $p_1=(\frac{2\Omega + \gamma}{3\alpha},\pi/2+k\pi)$, and hyperbolic saddles,  $p_2=(\frac{2\Omega - \gamma }{3\alpha},k\pi)$ ($k\in \mathbb{Z}$).
%\begin{equation}
%j_1 = \frac{\Omega + F_1/2}{2a}.
%\end{equation}
%for
%$\psi = \pi/2+k\pi$ ($k\in \mathbb{Z}$).
%and 
%\begin{equation}
%j_2 = \frac{\Omega - F_1/2}{2a},
%\end{equation}
%for $\psi = k\pi$. 
%Linear stability analysis shows that \((j_1,\pi/2)\) is a centre and \(p_2 = (j_2,0)\) is a hyperbolic saddle. 
Note that $0 < j_2 < j_1$. The homoclinic orbits $\Gamma_\pm= \left\{ (J_\pm(\tau), \psi_\pm(\tau)) \mid \tau \in \mathbb{R} \right\}$, 
that lie on the energy level of the saddle, $H_s = H(j_2,0) =  -\frac{3}{4}\alpha j_2^2$,
%\begin{equation}
%H_s = H(j_2,0) = -\frac{1}{4a}\left(\Omega - \frac{F_1}{2}\right)^2= -a j_2^2. \label{eq:Hs}
%\end{equation}
are provided by the implicit equation 
\begin{equation}
\dfrac{3}{4}\alpha J^2 - \Omega J + \frac{\gamma}{2}J\cos 2\psi + \frac{1}{3 \alpha}\left(\Omega - \frac{\gamma}{2}\right)^2 = 0. \label{eq:separatrix_implicit}
\end{equation}
Solving \eqref{eq:separatrix_implicit} for $J$ yields the two branches of the separatrix loop: 
\begin{equation}
J_{\pm}(\psi) = 
\frac{2\Omega -  \gamma \cos 2\psi \pm 2\sqrt{\gamma\sin^2\psi \left(2\Omega - \gamma\cos^2\psi\right)}}{ 3 \alpha} .
\label{eq:Jpm}
\end{equation} 

\begin{figure}[htb!]
\centering
\includegraphics[width=0.8\textwidth]{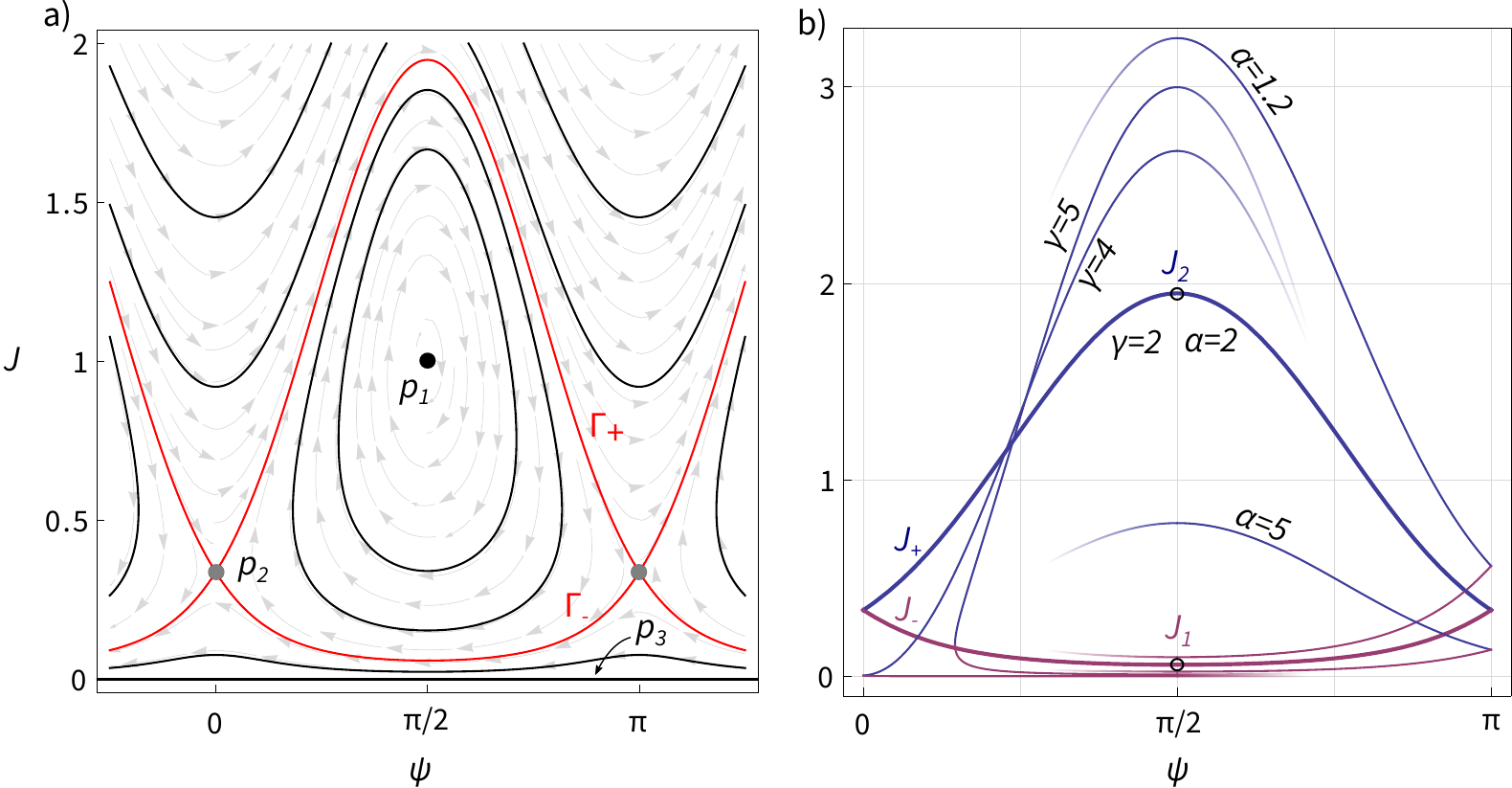}
\caption{Hamiltonian skeleton of the averaged system ($\mu=0$).
(a)~Phase portrait in the action-angle plane: level curves of $H$,
Eq.~\eqref{eq:Hamiltonian}, with the direction of the flow (grey arrows). $p_1$:
centre (phase state); $p_2$: saddles; $p_3$: rest state, $J=0$. The
separatrix (red) has an outer branch $\Gamma_{+}$
%= \left\{ (J_+(\tau), \psi_+(\tau)) \mid \tau \in \mathbb{R} \right\}$ 
and an inner branch
$\Gamma_{-}$. It separates the librations around $p_1$ from the
rotations above $\Gamma_{+}$ and below $\Gamma_{-}$. Energy Levels:
$H=\left\lbrace -10, -2, 1., 2.5, 5 \right\rbrace \times H_s$.
(b)~Separatrix branches $J_{+}(\psi)$ (blue) and $J_{-}(\psi)$
(purple), Eq.~\eqref{eq:Jpm}. Thick lines: $\gamma=2$, $\alpha=2$, $\Omega=2.01$. 
%Thin lines: $\gamma=4$ and $\gamma=5$ at $\alpha=2$; $\alpha=1.2$ and $\alpha=5$ at $\gamma=2$. 
Circles: turning points $J_{1,2}$,   
Eq.~\eqref{eq:turning}.}
\label{fig:hamiltonian}%Phase portrait of the Hamiltonian system and homoclinic branches. a) Energy Levels  $H=\left\lbrace -10, -2, 1., 2.5, 5 \right\rbrace \times H_s$. b) Variation of the homoclinic branches with the ampllitude of the fast excitation $\gamma$ and the Duffing nonlinearity.}
%\label{fig:PhasePortrait}
\end{figure} 
%The phase portrait of the system with $H(J,\psi)=H_s$  is shown in Fig.~\ref{fig:hamiltonian}.  
 % Figure 2
The Hamiltonian phase portrait of Fig.~\ref{fig:hamiltonian}(a)
organizes the dynamics and everything that follows. The separatrix, $H(J,\psi)=H_s$ (red curve), divides the plane
into three regions: librations around a phase state, inside the loop;
inner rotations around the rest state, below $\Gamma_{-}$; and outer
rotations, above $\Gamma_{+}$, which encircle both phase states and the
rest state. The two branches $J_{\pm}(\psi)$ 
%enclose the centre and 
meet at the saddle points  where the square root in Eq.~\eqref{eq:Jpm} vanishes. 
% $p1=(j_1,\pi/2+n \pi)$. 
%Alternatively, solving for $\phi$ in terms of $J$ gives the inverse branches
%\begin{equation}
%\phi_{\pm}(J) = 
%\pm \frac{1}{2}
%\arccos\!\left[
%\frac{2\Omega}{F_1} - \frac{2a}{F_1}J
%- \frac{(\Omega - F_1/2)^2}{2a F_1 J}
%\right] + n\pi, \qquad n \in \mathbb{Z}.
%\label{eq:phipm}
%\end{equation}

The homoclinic solution $J(\tau)$ is obtainable in closed form combining the two equations in \eqref{eq:systemHamiltonian} for the separatrix energy $H=H_s$, it gives  
\begin{equation}
\begin{aligned}
\dot J^2 = 3\alpha\,(J-j_2)^2\,\Bigl[\gamma J - \dfrac{3}{4}\alpha(J-j_2)^2\Bigr]. \label{eq:Jdot2_factored}
\end{aligned}
\end{equation} 
The motion on the separatrix is confined to the region where the factor in the brackets is non‑negative.
%:
%\begin{equation}
%f(J) = \gamma J - \dfrac{3}{4}\alpha(J-%j_2)^2 \ge 0.
%\label{eq:Jquadratic}
%\end{equation}
  Its roots, $0<J_1 < j_2 < J_2 $,  are  
\begin{equation}
J_{1,2} =j_2 +  \frac{2( \gamma \mp \sqrt{\gamma^2 + 3\alpha \gamma j_2})}{3\alpha}  , 
\label{eq:turning}
\end{equation}
or equivalently, using  $j_2=(2\Omega-\gamma)/(3\alpha)$,
\begin{equation}
J_{1,2}=
\frac{1}{3\alpha}
\left(\sqrt{2\Omega}\mp\sqrt{\gamma}\right)^2. 
\label{eq:turning_explicit}
\end{equation}
 They correspond to the inner ($J_1$) and outer ($J_2$) turning points of the separatrix loop (circles on the teardrop-shaped  line of Fig.~\ref{fig:hamiltonian}(b)).   
 The dependence of the separatrix geometry on the system parameters
is   apparent from Eq.~\eqref{eq:turning_explicit}. Increasing the detuning
$\Omega$ shifts both turning points towards larger actions and
increases their separation, whereas increasing the parametric
excitation strength $\gamma$ moves the inner turning point towards
smaller actions and the outer one towards larger actions, thereby
widening the separatrix. In contrast, increasing the Duffing
coefficient $\alpha$ reduces the action scale of the entire
structure. In particular, the separation between the turning points
is $J_2-J_1={4\sqrt{2\gamma\Omega}}/{3\alpha}$. 
%\begin{equation}
%J_2-J_1=\frac{4\sqrt{2\gamma\Omega}}{3\alpha}.
%\end{equation}
Thus, the separatrix provides a direct representation of how the
competition between detuning, parametric excitation and nonlinear
stiffness organizes the conservative phase-space dynamics.
%Existence of homoclinic orbits requires that the the square root of Eq.~\eqref{eq:Jpm}must be real. For $\gamma>0$, this is automatically achieved if the nontrivial saddle has positive $J$,  i.e. $\Omega>\gamma/2$ (see curves for $\gamma=4,5$ in Fig.~\ref{fig:hamiltonian}(b))).
The existence of the separatrix requires the radicand in
Eq.~\eqref{eq:Jpm} to be non-negative for all $\psi$. For
$\gamma>0$, this condition is satisfied when $\Omega\geq\gamma/2$.
A finite-action separatrix exists only if the nontrivial saddle has
positive action, which requires $\Omega>\gamma/2$. This condition
is satisfied for $\gamma=4$, but not for $\gamma=5$, as illustrated
in Fig.~\ref{fig:hamiltonian}(b).

The homoclinic orbits for the outer (+) and inner (-) branches are given by solving \eqref{eq:Jdot2_factored}, i.e.
\begin{equation}
%\dot J = \pm 2a\,|J-j_2|\,\sqrt{(J_2-J)(J-J_1)} .
\dfrac{dJ}{|J-j_2|\,\sqrt{(J_2-J)(J-J_1)}}=\pm \dfrac{3}{2}\alpha\,d\tau, 
\label{eq:Jdot_sqrt}
\end{equation}
that gives
\begin{equation}
J_{\pm}(\tau) = j_2 \pm \frac{2\gamma j_2}{\sqrt{\gamma^2 + 3\alpha \gamma j_2}\cosh(\kappa \tau) \mp \gamma},
\qquad \kappa =  \sqrt{3\alpha \gamma j_2}.
\label{eq:Jpmclosedform}
\end{equation}
At $\tau=0$ the equation recovers the turning point $J_{1,2}$.  
%$and \lim_{\tau \to \pm\infty} (J_\pm(\tau), \phi_\pm(\tau))  = (j_2, 0) = p_2.$
As $\tau\to\pm\infty$, $J_{\pm}(\tau)\to j_2$, the hyperbolic saddle. 
The  expressions \eqref{eq:Jpmclosedform} will serve in Sec.~\ref{sec:Melnikov} as the unperturbed orbits in the Melnikov calculation when including the damping and forcing perturbations.

The Hamiltonian phase portrait provides the conservative skeleton of the dynamics. For $\mu>0$, damping and secondary stiffness modulation perturb this structure and replace the stationary responses of the unperturbed system with modulation-driven trajectories. 
Their existence and stability are investigated next.

%In particular, the fixed points $p_i$ persist, for sufficiently small $\mu$, as periodic solutions $(J_i(\tau),\psi_i(\tau))$ of the averaged system. 
%Their existence and stability are investigated next.
%{\corr Damping and modulation modify the scenario of Fig.~\ref{fig:hamiltonian} and turn each family into attractors of its own. The forced librations are the tori that grow out of the phase states [Fig.~\ref{fig:overview}(b); blue in Figs.~\ref{fig:nu7} and~\ref{fig:nu35}]. The inner family gives the low-amplitude orbit born from the rest state (purple in Fig.~\ref{fig:nu35}). The outer family gives the large orbit of Fig.~\ref{fig:nu7} (red). Which of the three survives, and which becomes chaotic, depends on where the modulation frequency sits relative to the libration frequency $\lambda_1$. }

\section{Stability  of the  quasiperiodic motion}
\label{sec:stability}
For sufficiently small $\mu>0$, the averaged system \eqref{eq:JdotPhidot} possesses hyperbolic periodic solutions $(J_i(\tau),\psi_i(\tau))$, $i=1,2$, near the fixed points $p_i$. These hyperbolic periodic orbits of the averaged system correspond to normally hyperbolic invariant closed curves (1-tori) $T_i$, $i=1,2$, of the Poincaré map $\mathcal{P}_{\mu}$ (defined by sampling the original flow at the fast excitation period). The solutions originating from the centers describe forced librations around the parametric phase states, while the saddle persists as a periodic orbit of saddle type and continues to organize the global phase-space structure. 
%In the full oscillator, these periodic solutions of the averaged equations correspond to invariant two-tori supporting quasiperiodic motion. 
Equivalently, they correspond to normally hyperbolic invariant two-tori for the suspended flow of the full system. The trajectories on these invariant tori represent quasiperiodic dynamics.

An approximate expressions for the tori $T_{1,2}$ is determined using a perturbation approach. We assume the solution of the averaged system  \eqref{eq:JdotPhidot} in the form  
\begin{equation}
    J_i(\tau)=j_i+\mu\xi(\tau),\quad i=1,2,\qquad \psi_1(\tau)=\dfrac{k\pi}{2}+\mu\eta(\tau), \qquad \psi_2(\tau)=k\pi+\mu\eta(\tau),
\end{equation}
that gives
\begin{equation}
     \dot{\xi}= \mp 2\gamma j_i \eta- 2\delta \omega_\text{\tiny F}j_i -\delta_\text{\tiny NL}\omega_\text{\tiny F} j_i^2, \qquad 
      \dot{\eta}=\dfrac{3}{2}\alpha\xi+\beta\cos\nu_0 \tau
      \label{eq:eq_periodic}
\end{equation}
where the upper choice of sign refers to $i=1$. 
The solutions of \eqref{eq:eq_periodic} are 
\begin{equation}
     \xi(\tau)=\pm\dfrac{2\gamma\beta j_i}{\nu_0^2\mp\lambda_i^2}\cos \nu_0 \tau, \qquad
      \eta(\tau)=\dfrac{\nu_0\beta}{\nu_0^2\mp\lambda_i^2}\sin \nu_0 \tau\mp\dfrac{\omega_\text{\tiny F}}{2\gamma}\left(2\delta+\delta_\text{\tiny NL}j_i\right)
      \label{eq:sol_periodic}
\end{equation}
in which $\lambda_i^2=3\gamma\alpha j_i$, ($\lambda_{1,2}^2=\gamma(2\Omega\pm\gamma)$ is the natural frequency squared of the linearised system. 

\begin{figure*}[htb!]
\centering
\includegraphics[width=1.0\textwidth]{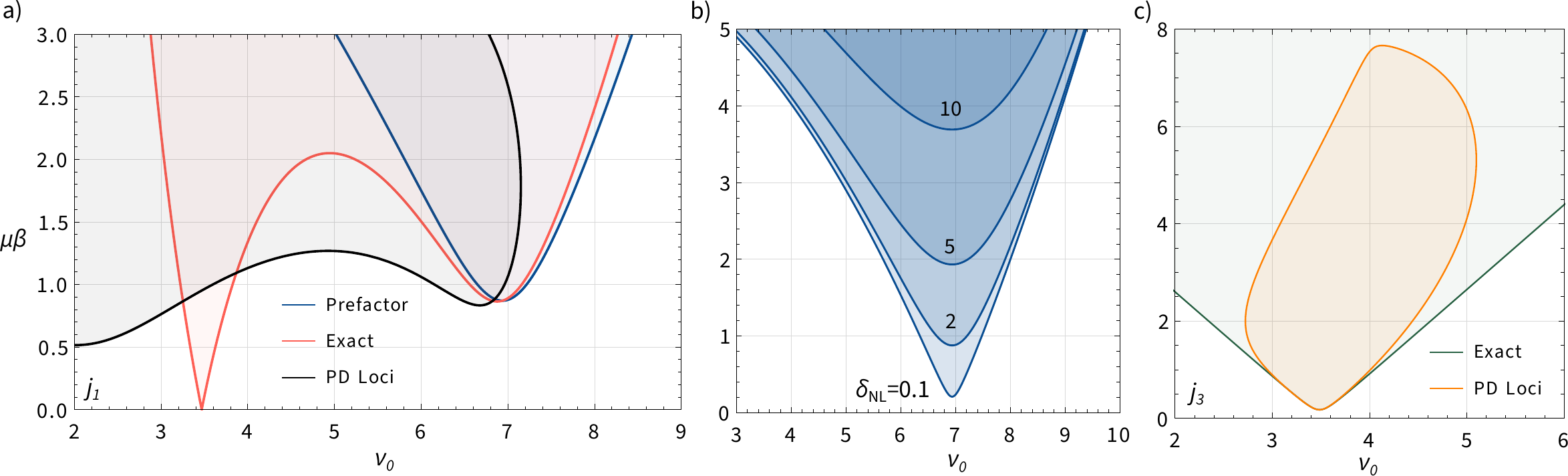}
\caption{
Stability of the forced librations in the $(\nu_0,\mu\beta)$
plane. (a)~Period-doubling instability of the torus around the phase
state $j_1$. Red: Eq.~\eqref{eq:instability}; its zero at $\nu_0=\lambda_1$ is
an artefact of the expansion. Blue:  
prefactor frozen   $4(\nu_0^2-\lambda_1^2)/{3\nu_0}\approx 2\lambda_1$. Black: period-doubling locus from
numerical continuation. Shading marks the unstable side of each curve.
(b)~Effect of the nonlinear damping $\delta_\text{\tiny NL}$ on the tongue's critical threshold. 
(c)~Instability of the rest state $j_3=0$. Green: Floquet threshold,
Eq.~\eqref{eq:inst_j3}. Orange: period-doubling locus from numerical
continuation, which closes into an island.  Parameters:  $\mu\beta=0.25$, $\delta=0.49505$, $\delta_{\text{\tiny NL}}=1.9802$ $\Omega=2.01$. }
%Instability region for the stable periodic orbit ($\mu\beta=0.25$, $\delta=0.49505$, $\delta_{\text{\tiny NL}}=1.9802$ $\Omega=2.01$). a) Instability threshold for $j_1$ computed via Eq.~\eqref{eq:instability} (red) and by using the approximated prefactor $4(\nu_0^2-\lambda_1^2)/{3\nu_0}\approx 2\lambda_1$ (blue). Black line is the numerical computation. b) Effect of the nonlinear damping $\delta_\text{\tiny NL}$ on the tongue's critical threshold. c)  Instability threshold for $j_3$ computed via Eq.~\eqref{eq:inst_j3} (green) and via numerical continuation (orange). }
\label{fig:Mathieu}
\end{figure*}

Since  $(J_2(\tau), \phi_2(\tau))$ is unstable, it is interesting to focus on the stability of the orbit $(J_1(\tau), \phi_1(\tau))$. The orbit is weakly perturbed as
\begin{equation}
 J= J_1(\tau)+\Delta J, \qquad \psi=\psi_1(\tau)+\Delta \psi,
 \label{eq:small_perturbation}
\end{equation}
that inserted in \eqref{eq:JdotPhidot} yields to  
\begin{equation}
\left\{
\begin{aligned}\dot{\Delta J}&=-\left[ 2\gamma j_1 +2\mu\gamma\xi\right]\Delta\psi -\mu\left[2   \omega_\text{\tiny F}\left(\delta+\delta_\text{\tiny NL} j_1\right)+2 \gamma\eta\right]\Delta J
\\
\dot{\Delta \psi}&=\dfrac{3}{2}\alpha\Delta J+2\mu\gamma\eta\Delta\psi
.
\end{aligned}
\right.
\label{eq:perturbed}
\end{equation}
At the leading order $\Delta\psi\approx - \dfrac{\dot{\Delta J}}{2\gamma j_1}$, thus Eq.~\eqref{eq:perturbed} becomes  
%\begin{equation}
%    \ddot{\Delta J}+\left[2\mu \omega_\text{\tiny F}\left(\delta+\delta_\text{\tiny NL} j_1\right)-\dfrac{\mu\dot{\xi}}{j_1}\right]\dot{\Delta J}+\left[\lambda_1^2+2\mu\gamma\dot{\eta}+3\mu\alpha\gamma\xi\right]\Delta J=0
%\end{equation}
\begin{equation}
    \ddot{\Delta J}+\left[2\mu\omega_\text{\tiny F}\left(\delta+\delta_\text{\tiny NL} j_1\right)+2\mu\beta\dfrac{ \gamma \nu_0}{ \nu_0^2-\lambda_1^2}\sin\nu_0 \tau\right]\dot{\Delta J}+\left[\lambda_1^2+2\mu\beta\dfrac{\gamma (\nu_0^2 +\lambda_1^2)  }{\nu_0^2-\lambda_1^2}\cos \nu_0 \tau\right]\Delta J=0, 
    \label{eq:Mathieu}
\end{equation}
which  is a damped Mathieu equation with both parametric damping and parametric stiffness.  
Equation \eqref{eq:Mathieu} is solved analytically introducing the transformation:
\begin{equation}
\Delta J = z_1(\tau)\cos\left(\frac{\nu_0}{2}\tau\right) -
z_2(\tau)\sin\left(\frac{\nu_0}{2}\tau\right),
\quad
\Delta\psi =
\frac{\nu_0}{4\gamma j_1}
\left[ z_1(\tau)\sin\left(\frac{\nu_0}{2}\tau\right) + z_2(\tau)\cos\left(\frac{\nu_0}{2}\tau\right)
\right].
\end{equation}
and applying the averaging method. This provides  
\begin{equation}
\left\{
\begin{aligned}
\dot{z}_1 &=- \mu\omega_\text{\tiny F}\left(\delta+\delta_\text{\tiny NL} j_1\right)z_1+\left(\dfrac{\left({\nu_0}/{2}\right)^2-\lambda^2}{\nu_0}+\dfrac{3\mu\beta\gamma\nu_0}{4(\nu_0^2-\lambda_1^2)}\right)z_2,\\ 
\dot{z}_2 &= \left(-\dfrac{\left({\nu_0}/{2}\right)^2-\lambda^2}{\nu_0}+\dfrac{3\mu\beta\gamma\nu_0}{4(\nu_0^2-\lambda_1^2)}\right)z_1- \mu\omega_\text{\tiny F}\left(\delta+\delta_\text{\tiny NL} j_1\right)z_2
.
\end{aligned}
\right.
\label{eq:avg_z12}
\end{equation}
%in which $\mu\sigma_1=\left({\nu_0}/{2}\right)^2-\lambda^2$. 
To establish the stability of the orbit, we analyse the eigenvalues of   \eqref{eq:avg_z12}. The orbit is unstable when the real part of an eigenvalue becomes positive, yielding the condition
\begin{equation}
(\mu\beta)^2
>
\frac{16(\nu_0^2-\lambda_1^2)^2}{9\gamma^2\nu_0^2}
\left\{
\left[\mu\omega_\text{\tiny F}(\delta+\delta_\text{\tiny NL} j_1)\right]^2
+
\left[\frac{(\nu_0/2)^2-\lambda_1^2}{\nu_0}\right]^2
\right\}.
\label{eq:instability}
\end{equation}
The instability threshold and the associated unstable region are depicted in red in Fig.~\ref{fig:Mathieu}(a). 
 The instability condition  \eqref{eq:instability} is   valid near the resonance center $\nu_0\approx 2\lambda_1$
 The spurious zero at $\nu_0\approx \lambda_1$ is outside the domain of validity of the averaging method (it corresponds to the primary linear resonance, where the periodic orbit amplitude diverges (see \eqref{eq:avg_z12}) and the perturbation expansion fails. 
 %it is a mathematical artifact corresponding to the primary linear resonance ($\nu_0\approx 2\lambda_1$)  where the averaging method breaks down entirely (the periodic orbit amplitude diverges).
The artifact can be removed by evaluating the prefactor on the right-hand side of Eq.~\eqref{eq:instability} at the centre of the resonance region (i.e., the centre of the domain over which the expansion is valid) and treating it as a constant. In particular,
$ {4(\nu_0^2-\lambda_1^2)}/{3\nu_0}
\approx 2\lambda_1$.
The numerical evaluation of Eq.~\eqref{eq:instability} and the corresponding result obtained using the constant prefactor are shown in Fig.~\ref{fig:Mathieu}(a). Replacing the frequency-dependent prefactor with a constant restores the characteristic Mathieu-tongue-like shape of the analytical prediction. 
An increase in the nonlinear damping shifts the instability tongue upwards, thereby increasing the activation threshold for period-doubling, as shown in Fig.~\ref{fig:Mathieu}(b).

The instability boundary is also determined numerically by continuing the loci of period-doubling bifurcations. The numerical results reveal a more intricate structure, which differs substantially from the classical Mathieu-type instability tongue (black curve/region in Fig.~\ref{fig:Mathieu}(a)). 
To further investigate this shape and the combined effects of parametric stiffness and damping modulation, we introduce a Liouville transformation 
\begin{equation} 
\Delta J(\tau)=
\exp\left[
-\mu\omega_{\text{\tiny F}}
\left(\delta+\delta_{\text{\tiny NL}}j_1\right)t
+
\frac{\mu\beta\gamma}{\nu_0^2-\lambda_1^2}
\cos(\nu_0 \tau)
\right]y(\tau). 
\label{eq:Liouville}
\end{equation} such that to  eliminate  the time-periodic damping coefficient in Eq.~ \eqref{eq:Mathieu} and recast it into   a into Hill form, i.e.
\begin{equation}
\begin{aligned}
\ddot y+\Bigg[
&\lambda_1^2
-\mu^2\omega_{\text{\tiny F}}^2
\left(\delta+\delta_{\text{\tiny NL}}j_1\right)^2
-\frac{\mu^2\beta^2\gamma^2\nu_0^2}
{2(\nu_0^2-\lambda_1^2)^2} +\frac{\mu\beta\gamma(\nu_0^2+2\lambda_1^2)}
{\nu_0^2-\lambda_1^2}\cos(\nu_0 \tau)
\\[2mm]
&-\frac{2\mu^2\beta\gamma\omega_{\text{\tiny F}}\nu_0
\left(\delta+\delta_{\text{\tiny NL}}j_1\right)}
{\nu_0^2-\lambda_1^2}\sin(\nu_0 \tau) +\frac{\mu^2\beta^2\gamma^2\nu_0^2}
{2(\nu_0^2-\lambda_1^2)^2}\cos(2\nu_0 \tau)
\Bigg]y=0.
\end{aligned}
\label{eq:Hill}
\end{equation}

Interestingly, the resulting effective periodic coefficient contains not only a fundamental harmonic at the parametric excitation frequency $\nu_0$, but also an additional second-harmonic stiffness modulation at $2\nu_0$.   
The additional second-harmonic contribution, proportional to $(\mu \beta)^2$ reveals a further mechanism through which parametric damping can contribute to the excitation of   the principal Mathieu resonance. In particular, when $\nu_0$ approaches $\lambda_1$, the second harmonic approaches twice the natural frequency of the unmodulated variational dynamics, potentially contributing to parametric instability. 
It is worth noting that the coefficients in Eq.~\eqref{eq:Hill} inherit the singularity at $\nu_0=\lambda_1$ already present in the periodic solution \eqref{eq:sol_periodic} used to derive Eq.~\eqref{eq:Mathieu}. Thus, although the Liouville transformation is exact, the resulting harmonic decomposition should be interpreted with caution in the vicinity of this resonance, where the perturbative expansion   ceases to be uniformly valid. 
%% maybe!!!! %% A systematic investigation of this resonant regime, based on a dedicated multiple-scales expansion of the reference solution and its associated  dynamics is left for future work. 

The \(j_3=0\) equilibrium also loses stability. The threshold can be
calculated via Floquet theory and is
\begin{equation}
\mu\beta=\frac{4\omega_{\tiny 0}}{\gamma}
\sqrt{(\mu\delta\omega_\text{\tiny F})^2
+\Bigl(\omega_{j_3}-\frac{\nu_0}{2}\Bigr)^2},
\label{eq:inst_j3}
\end{equation}
with $\omega_{j_3}=\sqrt{\Omega^2-\frac{\gamma^2}{4}} $, which at exact resonance \(\nu_0=2\omega_{\tiny 0}\) reduces to $
%\begin{equation}
\beta=\dfrac{4\omega_{j_3}\,\delta\omega_\text{\tiny F}}{\gamma}$.
%\end{equation}
Crossing the threshold, the dynamics passes directly from the stable
$j_3=0$ equilibrium to a torus that is invariant under the second
iterate $\mathcal{P}_{\mu}^2$ of the Poincar\'e map but not under
$\mathcal{P}_{\mu}$ itself.  Although Floquet theory correctly identifies the value of \(\mu\beta\)
at the activation, it does not predict the post-bifurcation dynamics,
which is governed by the cubic terms $\delta_\text{\tiny NL}$ and
$\alpha$. 
Comparison of the analytical and numerical  instability threshold is shown in Fig.~\ref{fig:Mathieu}(c).
A more in-depth analysis of the bifurcation scenario is presented in Sec.~\ref{sec:numerical}.

%, whose amplitude is quadratic in the parametric damping modulation and provides an additional mechanism for parametric excitation. Indded, when \(\nu_0\) approaches the natural frequency \(\lambda_1\), the second harmonic approaches twice the natural frequency, potentially contributing to the principal parametric instability. 

%However, the coefficients of the reduced variational equation become singular in this limit, reflecting the breakdown of the perturbative periodic-orbit expansion used in its derivation. Consequently, the harmonic decomposition identifies possible instability mechanisms but does not, by itself, provide a quantitatively valid description of the dynamics at this resonance. The transformation into Hill form reveals that parametric damping generates an additional second-harmonic contribution to the effective periodic coefficient, which may become relevant when the modulation frequency approaches the natural frequency of the linearised system. However, this frequency range also corresponds to a resonance in the perturbative construction of the reference periodic solution, as indicated by the singular dependence of the modulation coefficients on $(\nu_0^2-\lambda_1^2)$. Consequently, the present stability analysis is not uniformly valid in the vicinity of $(\nu_0=\lambda_1)$, and the instability features observed in this region cannot be quantitatively interpreted within the current approximation. 

{\corr

}

%perturbative description of the reference periodic solution is breaking down

%retain the Hill-equation transformation in the paper. It provides a useful structural interpretation of the combined parametric damping and stiffness modulation.

\section{Numerical analysis}
\label{sec:numerical}

\begin{figure}[htb!]
\centering
\includegraphics[width=1\textwidth]{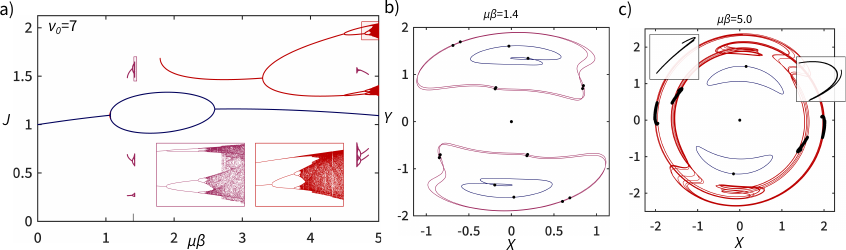}
\caption{Modulation at twice the libration frequency, $\nu_{\tiny 0}=7$.
(a)~Bifurcation diagram of the Poincar\'e map  $\mathcal{P}_\mu$ of Eq.~\eqref{eq:XY} versus $\mu\beta$. Blue: torus around a phase state, with a period-doubling
bubble for $1.06\lesssim\mu\beta\lesssim2.6$. Red: outer orbit, which
appears at $\mu\beta\approx1.8$, period-doubles at
$\mu\beta\approx3.3$ and becomes chaotic through a cascade (red
inset). Purple: rare attractors that exist in narrow windows near
$\mu\beta\approx1.35$ and $4.7$ (purple inset). (b)~Coexisting
attractors in the $(X,Y)$ plane at $\mu\beta=1.4$: the rest state (dot
at the origin), the period-doubled tori around the two phase states
(blue) and the rare attractors (purple). (c)~Same at $\mu\beta=5.0$: the rest state, the tori around the phase states (blue) and the chaotic attractor (red). Black dots: stroboscopic points. Insets in~(c): enlargements of the stroboscopic section of the chaotic attractor.}
\label{fig:nu7}
%Nonlinear dynamics for $\nu_0=7$. a) Numerical bifurcation diagram for the Poincar'e map $\mathcal{P}_\mu$ of Eq.~\eqref{eq:XY}}.  \label{fig:bif}
\end{figure}

 At $\nu_0=7$, the secondary modulation acts near twice the natural
libration frequency, while remaining far from the primary resonance.
The bifurcation diagram in Fig.~\ref{fig:nu7}(a) shows that the
forced-libration torus remains relatively small until it loses
stability through period doubling at $\mu\beta\approx1.06$.
The original period-one response regains stability at
$\mu\beta\approx2.6$, delimiting a period-doubling bubble. At fixed
modulation frequency, this instability interval corresponds to the
folded period-doubling boundary shown by the black curve in
Fig.~\ref{fig:Mathieu}(a). Beyond this bubble, the forced
librations persist up to the largest modulation depth considered.
The zero-amplitude state also remains stable throughout the
investigated range, since $\nu_0=7$ lies outside its own
period-doubling instability region (see the locus of Fig.~\ref{fig:Mathieu}(c)).

Additional attractors exist only within narrow parameter windows,
near $\mu\beta\approx1.35$ and $4.7$. These responses undergo
period-doubling cascades of their own, as indicated by the purple
branches in Fig.~\ref{fig:nu7}(a) and the corresponding trajectories
in Fig.~\ref{fig:nu7}(b).

Complex persisting dynamics develops
elsewhere. A separate outer orbit appears at $\mu\beta\approx1.8$ and
undergoes period doubling at $\mu\beta\approx3.3$. A subsequent
bifurcation cascade leads to chaos near $\mu\beta\approx4.6$.
At $\mu\beta=5$, the Poincaré map develops a bounded, folded structure with a pronounced Hénon-like appearance, c.f. Fig.~\ref{fig:nu7}(c). The repeated stretching and folding of trajectories produces a chaotic attractor that coexists with, and surrounds, the two phase-state tori and the stable zero-amplitude state. The Lyapunov exponents per map iteration, $\lambda_1^{(P)}=0.215$ and $\lambda_2^{(P)}=-1.004$, confirm dissipative chaos: the positive largest exponent indicates exponential sensitivity to initial conditions, while the negative second exponent reflects contraction along the other Lyapunov direction.

\begin{figure*}[htb!]
\centering
\includegraphics[width=1\textwidth]{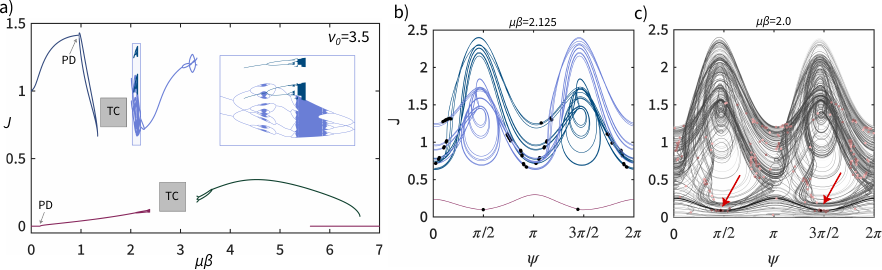}
\caption{Modulation at the libration frequency, $\nu_0=3.5$.
(a)~Bifurcation diagram: stroboscopic values of $J$ versus $\mu\beta$.
Upper family: the torus around a phase state (dark blue) period-doubles
(PD) at $\mu\beta\approx1.0$ and ends at $\mu\beta\approx1.35$; further
attractors (light blue) exist for $2.05\lesssim\mu\beta\lesssim2.2$,
enlarged in the inset, and for $2.3\lesssim\mu\beta\lesssim3.3$. Lower
family: the rest state period-doubles at $\mu\beta\approx0.2$ into a
low-amplitude orbit (purple) and is stable again for
$\mu\beta\gtrsim5.6$; a second low-amplitude orbit (green) exists for
$3.3\lesssim\mu\beta\lesssim6.6$. TC: intervals of transient chaos.
(b)~Coexisting attractors in the $(\psi,J)$ plane at $\mu\beta=2.125$:
two symmetry-related large orbits (light and dark blue) and the
low-amplitude period-2 orbit (purple). Black dots: stroboscopic points.
(c)~$\mu\beta=2.0$: a trajectory started near a phase state
wanders chaotically (grey; its stroboscopic points in shades of pink) before
settling on the low-amplitude period-2 orbit (black line; arrows mark
its two stroboscopic points).}
\label{fig:nu35}
\end{figure*}
At $\nu_0=3.5$, the modulation acts at the libration frequency, and the outcome is different in kind (Fig.~\ref{fig:nu35}). The forced libration is resonant: on the stroboscopic section in Fig.~\ref{fig:nu35}(a), the action rises from $1.0$ to about $1.4$ by $\mu\beta\approx1$. The torus period-doubles at $\mu\beta\approx1.0$, and the period-doubled torus disappears at $\mu\beta\approx1.35$. No attractor related to either phase state is found beyond this point. 
Indeed, for $1.35\lesssim\mu\beta\lesssim2.05$, a trajectory started near a phase state wanders erratically over both libration regions and finally settles on the low-amplitude orbit (Fig.~\ref{fig:nu35}(c)): the chaos is transient. Sustained chaos is found only within a narrow interval, $2.05\lesssim\mu\beta\lesssim2.2$, where large-amplitude trajectories visit both phase states. This attractor is illustrated in the inset of Fig.~\ref{fig:nu35}(a) and in Fig.~\ref{fig:nu35}(b). At $\mu\beta=2.15$, the Poincaré map exhibits dissipative chaos, with Lyapunov exponents per map iteration $\lambda_1^{(P)}=0.176$ and $\lambda_2^{(P)}=-1.330$. The positive largest exponent indicates exponential stretching, while the negative second exponent reflects contraction in the transverse direction, confirming the chaotic nature of the bounded attractor.

%Indeed, for $1.35\lesssim\mu\beta\lesssim2.05$, a trajectory started near a phase state wanders erratically over both libration regions and finally settles on the low-amplitude orbit (Fig.~\ref{fig:nu35}(c)): the chaos is transient. Sustained chaos appears only within a window, $2.05\lesssim\mu\beta\lesssim2.2$, on large orbits that visit both phase states. This is illustrated in the inset of Fig.~\ref{fig:nu35}(a) and in Fig.~\ref{fig:nu35}(b). 
%The system exhibits window of dissipative chaotic character, e.g. at $\mu\beta=2.15$, the Poincaré map has a positive largest Lyapunov exponent and a negative second exponent, $\lambda_1^{(P)}=0.176$ and $\lambda_2^{(P)}=-1.330$, respectively. These values indicate exponential stretching accompanied by strong contraction in the transverse direction, consistent with the formation of a bounded chaotic attractor.

The rest state is also affected by modulation close to the libration frequency, because $\nu_0=3.5$ also lies at the tip of its instability island (Fig.~\ref{fig:Mathieu}(c)). It period-doubles at $\mu\beta\approx0.2$ into the low-amplitude orbit and recovers at $\mu\beta\approx5.6$. Among the attractors found, only low-amplitude motion remains beyond $\mu\beta\approx3.4$, and only the rest state beyond $\mu\beta\approx6.6$. A slow tone tuned to the libration frequency therefore quenches the parametric oscillation, whereas the same tone at twice that frequency leaves it in place.

The numerical results reveal that sustained chaos emerges only at modulation amplitudes considerably larger than those at which the invariant tori lose stability. To understand the origin of this complex dynamics, and in particular the role played by the homoclinic separatrix of the unperturbed Hamiltonian system, the following section examines the conditions under which the secondary modulation induces intersections of the stable and unstable manifolds of the saddle.

\section{Melnikov's method}
\label{sec:Melnikov}
To analytically investigate the onset of homoclinic intersections, consider Eq.~\eqref{eq:JdotPhidot} in the form 
\begin{equation}
\dot{\mathbf{x}} = \mathbf{f}(\mathbf{x}) + \mu \mathbf{g}(\mathbf{x},t), \qquad \mathbf{x}  = \left[J \quad \psi\right]^\text{T}%\begin{pmatrix} J   \\[4pt]  \psi \end{pmatrix}\label{eq:Mdef}
\end{equation} 
with $\mu$ a bookkeeping parameter governing the perturbation introduced by damping and slow modulation in   the Hamiltonian system, i.e.  
\begin{equation}
\mathbf{f}(\mathbf{x})  = \begin{pmatrix} \gamma J \sin 2\psi \\[4pt] -\Omega + \dfrac{3}{2}\alpha J + \dfrac{\gamma}{2}\cos 2\psi \end{pmatrix},   \qquad
\mathbf{g}(\mathbf{x},t)  = \begin{pmatrix} -2\delta\omega_\text{\tiny F} J - \delta_\text{\tiny NL}\omega_\text{\tiny F} J^2 \\[4pt] \beta \cos(\nu_0 \tau) \end{pmatrix}.
\label{eq:component}
\end{equation}

The Melnikov function is then defined by the formula \footnote{  the wedged product is defined by $\mathbf{a}\wedge\mathbf{b}=a_1 b_2-a_2 b_1$}
\begin{equation}
M(t_0) = \int_{-\infty}^{\infty} \mathbf{f}(\mathbf{q}_0(t)) \wedge \mathbf{g}(\mathbf{q}_0(t), \tau+\tau_0)\, dt.
\label{eq:Mdef}
\end{equation}

\begin{figure}[htb!]
\centering
\includegraphics[width=1\textwidth]{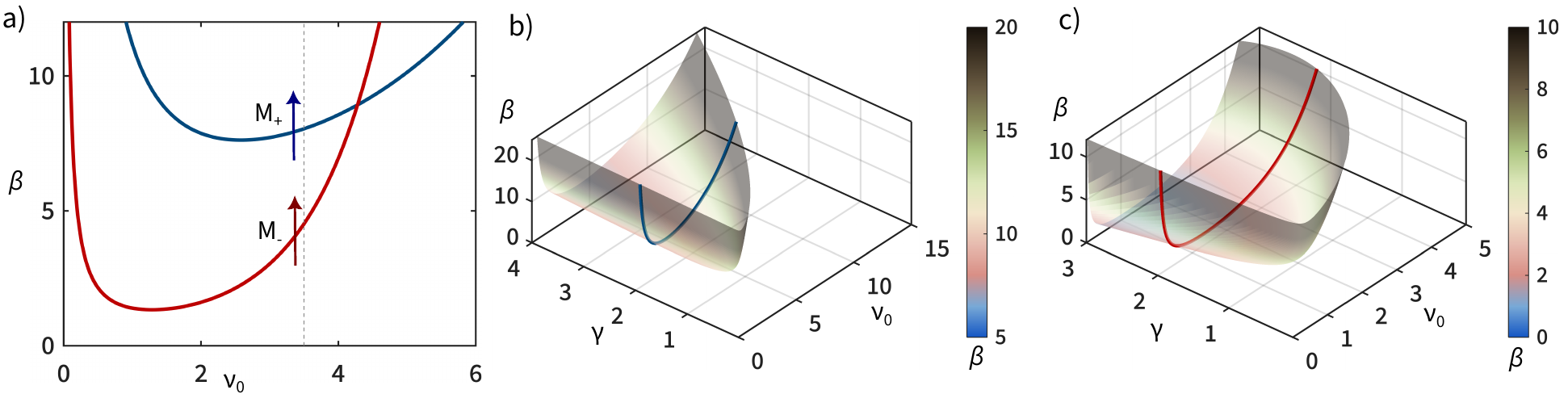}
\caption{Melnikov thresholds for transverse homoclinic intersections.
(a)~Critical modulation amplitude $\beta$
versus $\nu_0$ for $\gamma=2$ (, Eq.~\eqref{eq:betacr}), for the two branches of the separatrix
(red: $M_{-}$; blue: $M_{+}$). The dashed line marks $\nu_0=3.5$,
%the frequency of Fig.~\ref{fig:manifolds}; 
the arrows show the two
thresholds being crossed as $\beta$ increases. (b)~Threshold $M_{+}$ as
a function of $\nu_0$ and of the pump amplitude $\gamma$; the blue line
is the section $\gamma=2$ shown in~(a). (c)~Same for $M_{-}$; red line:
$\gamma=2$. 
%\todo{Other parameters: \ldots}.
}
\label{fig:melnikov} 
\end{figure}

By applying Eq.~\eqref{eq:Mdef} to Eq.~\eqref{eq:component}, the Melnikov function for $\Gamma_\pm$ become 
\begin{equation} 
\begin{aligned}
M_{\pm}(t_0) = \int_{-\infty}^{\infty} \Bigg\{ &
\left[ \gamma J_{\pm}(\tau) \sin(2\psi_{\pm}(\tau)) \right] \cdot \beta \cos(\nu_0(\tau+\tau_0)) \\
& - \left[ -\Omega + \dfrac{3}{2}\alpha J_{\pm}(\tau) + \dfrac{\gamma}{2}\cos(2\psi_{\pm}(\tau)) \right]  \left[ -2\delta \omega_\text{\tiny F} J_{\pm}(\tau) - \delta_\text{\tiny NL} \omega_\text{\tiny F} J_{\pm}^2(\tau) \right] \Bigg\} \, dt 
\end{aligned} 
\end{equation}
The integral is evaluated by employing  the method of residue \cite{Yagasaki1990}. For the outer ($+$) and inner ($-$) homoclinic branches it gives  
\begin{equation}
M_{\pm}(t_0)
=\pm
\beta\,\nu_0\,\dfrac{4\pi}{3\alpha}
\,
\dfrac{\sinh\!\left(\dfrac{\nu_0}{\kappa}\Theta_{\pm}\right)}
{\sinh\!\left(\dfrac{\pi\nu_0}{\kappa}\right)}\sin(\nu_0 \tau_0)
+
K_{\pm},
\label{eq:Mpm}
\end{equation}

with
\begin{equation}
\begin{aligned}
K_{\pm}
&=
2\omega_\text{\tiny F}
\Big[
\pm j_2\left(2\delta+\delta_\text{\tiny NL}j_2\right)\Theta_{\pm} 
+
\frac{\left(2\delta+3j_2\delta_\text{\tiny NL}\right)}
{3\alpha}
\left(\kappa\pm\gamma\Theta_{\pm}\right) +
\frac{\delta_\text{\tiny NL}\gamma}{3\alpha^{2}}
\left[
\pm\Theta_{\pm}\left(\gamma+\alpha j_2\right)
+\kappa
\right]
\Big],\\
\kappa&=\sqrt{3\alpha\gamma j_2},
\qquad
\Theta_{\pm}
=
\arccos\!\left(
\mp\frac{\gamma}{\sqrt{\gamma^{2}+3\alpha\gamma j_2}}
\right).
\label{eq:Theta_pm}
\end{aligned}
\end{equation}

\begin{equation} 
\beta_{\pm}^{\mathrm{cr}}(\nu_0)
=
\frac{3\alpha\left|K_{\pm}\right|}{4\pi}
\;
\frac{\sinh\!\left(\dfrac{\pi\nu_0}{\kappa}\right)}
{\nu_0\,
\left|
\sinh\!\left(\dfrac{\nu_0}{\kappa}\Theta_{\pm}\right)
\right|} 
\label{eq:betacr}
\end{equation}
The threshold curves given by Eq.~\eqref{eq:betacr} are plotted in Fig.~\ref{fig:melnikov}. 
The critical modulation amplitude as a function of $\nu_0$ is shown in panle a for the two separatrix branches. Both curves exhibit a non-monotonic dependence on the modulation frequency, with a minimum corresponding to the most susceptible frequency range for homoclinic intersections. The threshold for the $M_-$ branch (red) is consistently lower than that for the $M_+$ branch (blue), indicating the first transverse intersections. The dashed line marks $\nu_0=3.5$, and the arrows indicate the successive crossings of the two thresholds as $\beta$ is increased. Panels (b) and (c) show the corresponding threshold surfaces as functions of $\nu_0$ and $\gamma$ for the $M_+$ and $M_-$ branches, respectively, with the curves at $\gamma=2$ recovering the results in panel (a).
In both panels, the threshold surfaces terminate at the boundary $\gamma=2\Omega$, where the saddle equilibrium approaches the
zero-action state and the finite-action homoclinic orbit ceases to exist, consistently with the condition $\gamma<2\Omega$ for the
existence of the homoclinic separatrix.

%with $\theta_0 = \arccos\!\left(\frac{F_1}{\sqrt{\Delta}}\right)$
%       = \arctan\!\left(\frac{\alpha}{F_1}\right).$

\begin{figure}[H]
\centering
\includegraphics[width=0.7\textwidth]{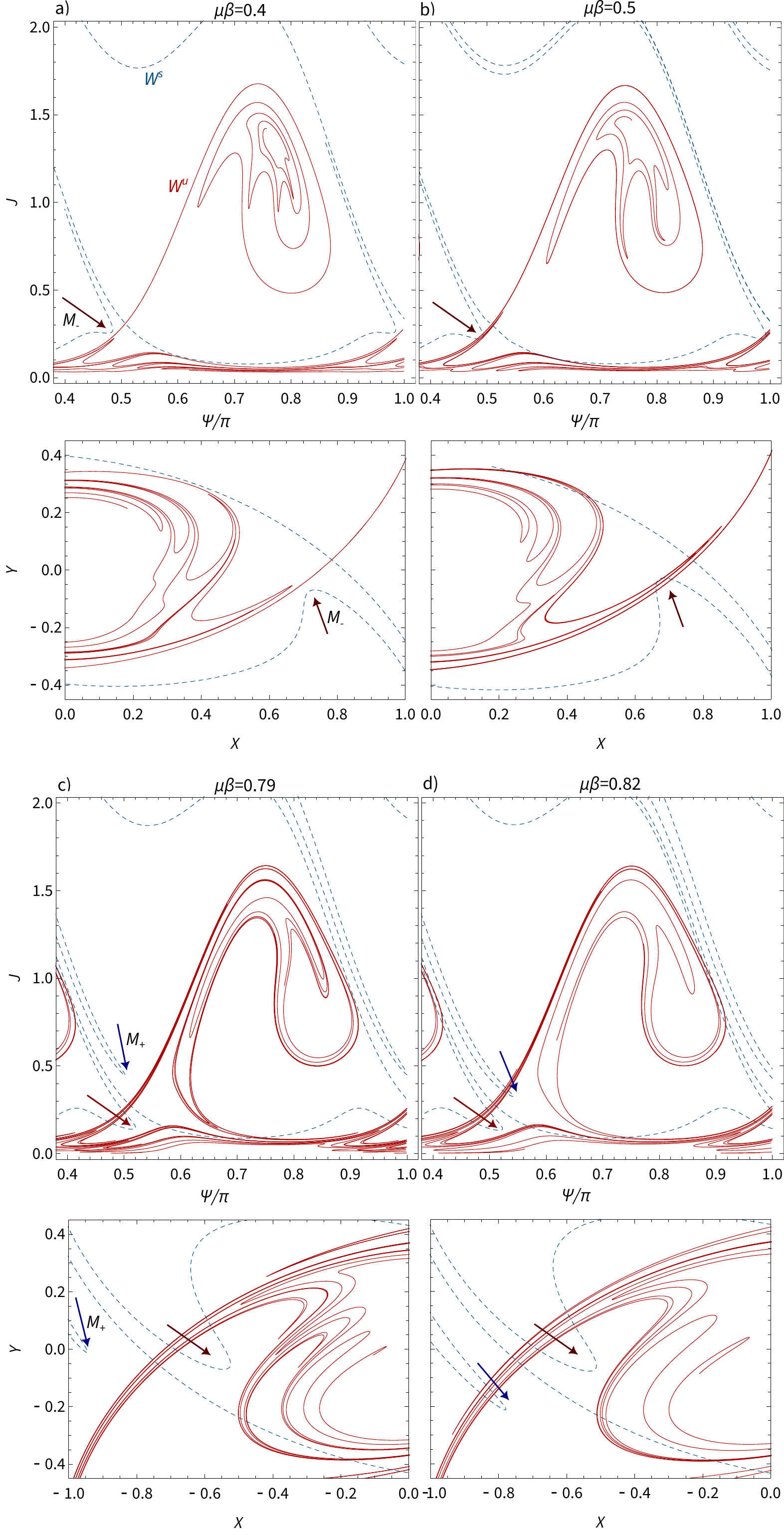}
\caption{Stable ($W^{s}$, dashed blue) and unstable ($W^{u}$, solid
red) manifolds of the saddle-type orbit of the $\mathcal{P}_\mu$ of Eq.~\eqref{eq:XY}, for $\nu_0=3.5$, on either side of the two
tangencies predicted in Fig.~\ref{fig:melnikov}(a). Upper panels:
$(\psi,J)$ plane; lower panels: enlargements in the $(X,Y)$ plane.
(a)~$\mu\beta=0.4$ and (b)~$\mu\beta=0.5$: before and after the first
tangency ($M_{-}$, dark-red arrows; predicted at $\mu\beta=0.4522$).
(c)~$\mu\beta=0.79$ and (d)~$\mu\beta=0.82$: before and after the
second tangency ($M_{+}$, blue arrows; predicted at
$\mu\beta=0.8043$). The manifolds are computed with the algorithm in 
Ref.~\cite{Parker1987}.} \label{fig:manifolds}
\end{figure}

\begin{figure}[H]
\centering
\includegraphics[width=0.5\textwidth]{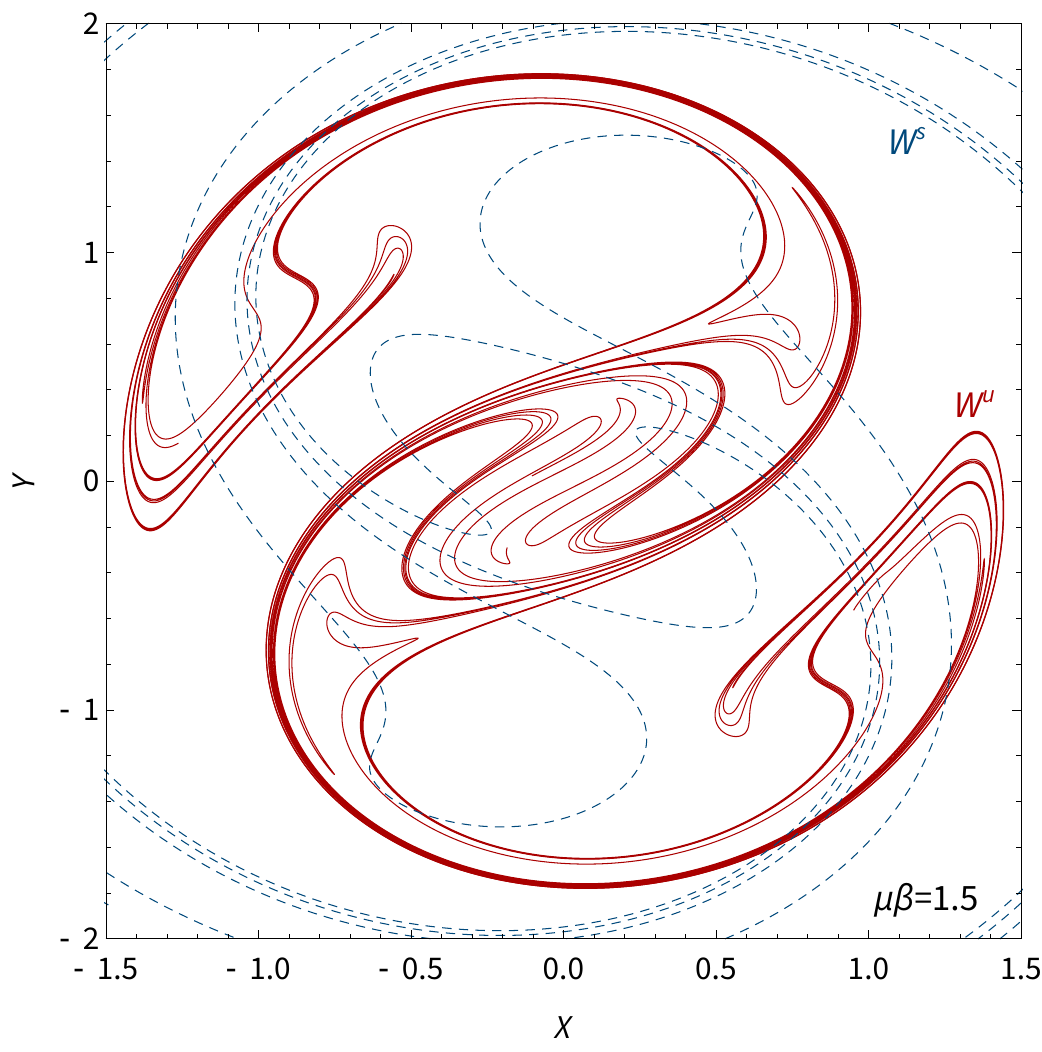}
\caption{Homoclinic tangle at $\mu\beta=1.5$, $\nu_0=3.5$  in the
$(X,Y)$ plane: stable ($W^{s}$, dashed blue) and unstable ($W^{u}$,
solid red) manifolds of the saddle-type orbit. The picture is invariant
under $(X,Y)\to(-X,-Y)$, the symmetry that exchanges the two phase
states.}
\label{fig:tangle}
\end{figure}

Figure~\ref{fig:melnikov}(a) places these results in the adiabatic and
nonadiabatic picture. Both thresholds of Eq.~\eqref{eq:betacr} diverge
as $1/\nu_0$ when $\nu_0$ tends to zero, because a slowly varying
detuning displaces the separatrix without splitting it. Both grow
exponentially at large $\nu_0$, where the modulation averages out. The
homoclinic tangle is therefore a nonadiabatic effect, confined to
modulation frequencies of the order of $\lambda_2$, the rate of the
saddle, which is itself of the order of the libration frequency.   
  In particular, the Melnikov analysis for $\nu_0=3.5$ it  predicts tangencies at $\mu\beta=0.452$ and $0.804$ (Fig.~\ref{fig:melnikov}(a)). 
The computed manifolds confirm both. 
%: the lower manifolds first touch between $\mu\beta=0.4$ and $0.5$, and the upper ones between $0.79$ and $0.82$. 
Figure~\ref{fig:manifolds} displays plots of the stable and unstable manifolds, $W^s$ (blue) and $W^u$ (red), which are computed using the Poincar\'e map $\mathcal{P}_\mu$ of Eq.~\eqref{eq:XY}.   
 The lower stable and unstable manifolds first intersect at approximately  for $0.4<\mu\beta<0.5$  (panels a and b)
while the upper manifolds intersect for $0.79<\mu\beta<0.82$.   
Both tangencies precede the period doubling of the torus and its disappearance. For $0.45<\mu\beta<1.35$ a horseshoe therefore coexists with regular attractors. It leaves no trace in the bifurcation diagram, but it prepares what follows.
The tangle of Fig.~\ref{fig:tangle} connects the libration regions with
the neighbourhood of the rest state. Once the torus is lost, the tangle
supplies both the chaotic transient and the route to the low-amplitude
orbit seen in Fig.~\ref{fig:nu35}(c). 
%These results are coherent with the theoretical estimation (See Fig.\ref{fig:melnikov}(a)) provided by the Melnikov method in Sec.~\ref{sec:Melnikov} that predicts transverse intersections for $\nu_0=3.5$ at $\mu\beta=0.4522$ and $0.8043$, respectively.
%As shown in panles c the homoclinic bifurcation constructs a horseshoe map. However, stable periodic points exist to which almost all orbits converge, meaning chaotic motions are not observed until $\mu\beta$ reaches a significantly higher value. 

 %{\corr
 %The computed manifolds confirm both: the lower manifolds first touch between $\mu\beta=0.4$ and $0.5$, and the upper ones between $0.79$ and
%$0.82$ (Fig.~\ref{fig:manifolds}). Both tangencies precede the period doubling of the torus and its disappearance. For $0.45<\mu\beta<1.35$ a horseshoe therefore coexists with regular attractors. It leaves no trace in the bifurcation diagram, but it prepares what follows.

%The two cases are compared in Table~\ref{tab:comparison}.
%\begin{table}[t]
%\centering
%\caption{The two nonadiabatic resonances compared. Values of $\mu\beta$ preceded by $\approx$ are read from Figs.~\ref{fig:nu7} and~\ref{fig:nu35}.}
%\label{tab:comparison}
%\renewcommand{\arraystretch}{1.25}
%\begin{tabular}{p{0.23\textwidth}p{0.35\textwidth}p{0.33\textwidth}}
%\hline
% & $\nu_0=7$ ($2\lambda_1$) & $\nu_0=3.5$ ($\lambda_1$) \\
%\hline
%Rest state & Stable throughout & Period-doubles at $\approx0.2$; stable again above $\approx5.6$ \\
%Torus around a phase state & Period-doubling bubble, $\approx1.06$ to $\approx2.6$, then stable   up to $5$& Period-doubles at $\approx1.0$; disappears at $\approx1.35$ \\
%Homoclinic tangencies & {\corr compute from Eq.~\eqref{eq:betacr}} & $0.452$ and $0.804$ \\
%Chaos & On the outer orbit, through a cascade near $\approx4.6$; narrow  window near $\approx1.4$
% & Transient from $\approx1.35$; sustained for   $2.05\lesssim\mu\beta\lesssim2.2$ \\
%Parametric oscillation & Survives & Quenched \\
%\hline
%\end{tabular}
%\end{table}
%}

%Figure 10 illustrates a bifurcation diagram for the Poincaré map \(\tilde{P}\)... For small \(\mu\beta\), there exist... (text cuts off)

\section{Conclusions}
 \label{sec:conclusions}
This work  investigates the dynamics of a parametrically excited
Duffing oscillator subject to a secondary modulation of its stiffness.
Although the modulation is slow compared with the primary excitation,
its timescale can become comparable to the intrinsic libration timescale
of the parametric phase states, leading to a breakdown of the adiabatic
response. 
%The resulting dynamics have been analysed by combining averaging, Floquet theory, numerical continuation, and Melnikov's method.
%The averaged equations reveal how the two phase states of the parametrically excited oscillator are affected by the secondary modulation. In the absence of modulation, the phase portrait has a Hamiltonian structure, with  centres surrounded by a homoclinic separatrix connecting saddle points. 
%For nonzero secondary modulation, the responses originating from the centres persist as periodic solutions of the averaged system, 
For nonzero secondary modulation the dynamics evolves through  invariant two-tori of the full flow. These solutions describe modulation-driven librations around the parametric phase states. Their stability is governed by a damped Mathieu equation, in which the secondary modulation contributes to both the effective stiffness and damping.
This formulation provides analytical access to the period-doubling
instability boundaries and also identifies a distinct instability
region associated with the zero-amplitude state.

When the modulation approaches the natural libration frequency, the forced librations are resonantly amplified, eventually leading to the loss of the phase-state invariant tori. The resulting dynamics include transient chaotic motion and a finite parameter interval of sustained chaos, before the parametric oscillation is ultimately quenched.
At twice the libration frequency, the response is markedly different: the phase-state tori remain stable over a much broader range of modulation amplitudes, apart from a bounded period-doubling instability region. In this case, sustained chaos develops on a separate outer attractor and coexists with the phase-state tori and the stable zero-amplitude state. Thus, the same secondary modulation can either suppress the parametric oscillation or promote complex dynamics while leaving the phase states intact, depending on its resonance condition.

 Via Melnikov's method the global mechanism underlying the onset of complex dynamics is investigated.  
%been investigated through Melnikov's method. The closed-formhomoclinic orbits of the unperturbed system make it possible to derive 
The explicit critical modulation amplitudes for transverse
intersections of the stable and unstable manifolds, exhibit
a non-monotonic dependence on the modulation frequency. They diverge in both the quasi-static and high-frequency limits, and attain minimum values at intermediate frequencies set by the characteristic dynamics near the saddle. Numerical computations of the invariant
manifolds corroborate these predictions. Importantly, the
homoclinic intersections occur at modulation amplitudes well below
those required for the development of sustained chaotic attractors. They therefore signal the onset of a global reorganization of the phase-space structure, rather than the immediate onset of sustained chaos.

Overall, these results show that the relative timescales of the modulation and the intrinsic phase dynamics determine whether the phase states can follow the forcing, undergo resonant amplification, or lose stability. The combined description based on invariant-torus stability and homoclinic manifold dynamics provides a framework for understanding these distinct routes to complex behaviour in parametrically driven nonlinear oscillators.

%%%%%%%%%% Insert bibliography here %%%%%%%%%%%%%%
 \bibliography{biblio}

@incollection{LifshitzCross2008,
  author    = {Lifshitz, Ron and Cross, M. C.},
  title     = {Nonlinear dynamics of nanomechanical and micromechanical resonators},
  booktitle = {Reviews of Nonlinear Dynamics and Complexity},
  publisher = {John Wiley \& Sons},
  year      = {2008},
  chapter   = {1},
  pages     = {1--52}
}

@article{Steeneken2021,
	author={Steeneken, Peter G and Dolleman, Robin Joey and Davidovikj, Dejan and Alijani, Farbod and van der Zant, Herre S J},
	title={Dynamics of 2D Material Membranes},
	journal={2D Materials},
	volume = {8},
	number = {4},
	pages = {042001},
	OPTurl={http://iopscience.iop.org/article/10.1088/2053-1583/ac152c},
	doi={10.1088/2053-1583/ac152c},
	year={2021}
}

@Article{Belardinelli2025,
author={Belardinelli, P.
and Yang, W.
and Bachtold, A.
and Dykman, M. I.
and Alijani, F.},
title={Hidden Vibrational Bistability Revealed by Intrinsic Fluctuations of a Carbon Nanotube},
journal={Nano Letters},
year={2025},
OPTmonth={May},
OPTday={28},
publisher={American Chemical Society},
volume={25},
number={21},
pages={8443-8449},
issn={1530-6984},
doi={10.1021/acs.nanolett.4c06618},
OPTurl={https://doi.org/10.1021/acs.nanolett.4c06618}
}

@ARTICLE{Parker1987,
  author={Parker, T.S. and Chua, L.O.},
  journal={Proceedings of the IEEE}, 
  title={Chaos: A tutorial for engineers}, 
  year={1987},
  volume={75},
  number={8},
  pages={982-1008},
  doi={10.1109/PROC.1987.13845}}

@misc{wattjes2026,
      title={Nonadiabatic dynamics under slow parametric modulation in nonlinear graphene resonators}, 
      author={Chris F. D. Wattjes and Haemin Lim and Pierpaolo Belardinelli and John E. Sader and Peter G. Steeneken and Farbod Alijani},
      year={2026},
      opteprint={2610.19816},
      archivePrefix={arXiv},
      primaryClass={cond-mat.mes-hall},
      url={https://arxiv.org/abs/2610.19816}, 
}

@article{Ardister2024,
    author = {Ardister, Jamal and Afzali, Fatemeh and Feeny, Brian F.},
    title = {Primary Resonance in a Weakly Forced Oscillator With Both Parametric Damping and Stiffness},
    journal = {Journal of Vibration and Acoustics},
    volume = {146},
    number = {1},
    pages = {011003},
    year = {2024},
    month = {04},
    issn = {1048-9002},
    doi = {10.1115/1.4065037},
    OPTurl = {https://doi.org/10.1115/1.4065037},
    OPTeprint = {https://asmedigitalcollection.asme.org/vibrationacoustics/article-pdf/146/1/011003/7325560/vib_146_1_011003.pdf},
}

@misc{margiani2026activatedswitchingcoexistinglimit,
      title={Activated switching between coexisting limit cycles}, 
      author={Gabriel Margiani and Orjan Ameye and Oded Zilberberg and Alexander Eichler},
      year={2026},
      eprint={2608.19060},
      archivePrefix={arXiv},
      primaryClass={math.DS},
      url={https://arxiv.org/abs/2608.19060}, 
}

@article{Faraday1831,
    author = {Faraday, Michael},
    title = {XVII. On a peculiar class of acoustical figures; and on certain forms assumed by groups of particles upon vibrating elastic surfaces},
    journal = {Philosophical Transactions of the Royal Society of London},
    number = {121},
    pages = {299-340},
    year = {1831},
    month = {12},
    issn = {0261-0523},
    doi = {10.1098/rstl.1831.0018},
    OPTurl = {https://doi.org/10.1098/rstl.1831.0018},
    OPTeprint = {https://royalsocietypublishing.org/rstl/article-pdf/doi/10.1098/rstl.1831.0018/1453165/rstl.1831.0018.pdf},
}

@article{Catalini2025,
  title = {Slow and fast topological dynamical phase transitions in a Duffing resonator driven by two detuned tones},
  author = {Catalini, Letizia and del Pino, Javier and Kumar, Soumya S. and Dumont, Vincent and Margiani, Gabriel and Zilberberg, Oded and Eichler, Alexander},
  journal = {Phys. Rev. Res.},
  volume = {7},
  issue = {3},
  pages = {033058},
  numpages = {6},
  year = {2025},
  OPTmonth = {Jul},
  publisher = {American Physical Society},
  doi = {10.1103/y9gq-yjxy},
  OPTurl = {https://link.aps.org/doi/10.1103/y9gq-yjxy}
}

@article{Kumar2026,
  title = {Dynamical phase transitions across slow and fast regimes in a two-tone driven Duffing resonator},
  author = {Kumar, Soumya S. and del Pino, Javier and Catalini, Letizia and Eichler, Alexander and Zilberberg, Oded},
  journal = {Phys. Rev. Res.},
  volume = {8},
  issue = {2},
  pages = {023298},
  numpages = {15},
  year = {2026},
  OPTmonth = {Jun},
  publisher = {American Physical Society},
  doi = {10.1103/d3b6-lnw6},
  OPTurl = {https://link.aps.org/doi/10.1103/d3b6-lnw6}
}

@article{Rugar1991,
  title = {Mechanical parametric amplification and thermomechanical noise squeezing},
  author = {Rugar, D. and Gr{\"u}tter, P.},
  journal = {Phys. Rev. Lett.},
  volume = {67},
  issue = {6},
  pages = {699--702},
  numpages = {0},
  year = {1991},
  month = {Aug},
  publisher = {American Physical Society},
  doi = {10.1103/PhysRevLett.67.699},
  OPTurl = {https://link.aps.org/doi/10.1103/PhysRevLett.67.699}
}

@Article{Mahboob2008,
author={Mahboob, I.
and Yamaguchi, H.},
title={Bit storage and bit flip operations in an electromechanical oscillator},
journal={Nature Nanotechnology},
year={2008},
month={May},
day={01},
volume={3},
number={5},
pages={275-279},
issn={1748-3395},
doi={10.1038/nnano.2008.84},
OPTurl={https://doi.org/10.1038/nnano.2008.84}
}

@article{Mahboob2016,
author = {Imran Mahboob  and Hajime Okamoto  and Hiroshi Yamaguchi },
title = {An electromechanical Ising Hamiltonian},
journal = {Science Advances},
volume = {2},
number = {6},
pages = {e1600236},
year = {2016},
doi = {10.1126/sciadv.1600236},
OPTURL = {https://www.science.org/doi/abs/10.1126/sciadv.1600236},
OPTeprint = {https://www.science.org/doi/pdf/10.1126/sciadv.1600236}}

@article{Margiani2025,
  title = {Three Strongly Coupled Kerr Parametric Oscillators Forming a Boltzmann Machine},
  author = {Margiani, Gabriel and Ameye, Orjan and Zilberberg, Oded and Eichler, Alexander},
  journal = {Phys. Rev. Lett.},
  volume = {135},
  issue = {9},
  pages = {097201},
  numpages = {7},
  year = {2025},
  OPTmonth = {Aug},
  publisher = {American Physical Society},
  doi = {10.1103/npp4-b1xb},
  OPTurl = {https://link.aps.org/doi/10.1103/npp4-b1xb}
}

@Article{Grimm2020,
author={Grimm, A.
and Frattini, N. E.
and Puri, S.
and Mundhada, S. O.
and Touzard, S.
and Mirrahimi, M.
and Girvin, S. M.
and Shankar, S.
and Devoret, M. H.},
title={Stabilization and operation of a Kerr-cat qubit},
journal={Nature},
year={2020},
OPTmonth={Aug},
day={01},
volume={584},
number={7820},
pages={205-209},
issn={1476-4687},
doi={10.1038/s41586-020-2587-z},
opturl={https://doi.org/10.1038/s41586-020-2587-z}
}

@article{ZounesRand1998,
author = {Zounes, Randolph S. and Rand, Richard H.},
title = {Transition Curves for the Quasi-Periodic Mathieu Equation},
journal = {SIAM Journal on Applied Mathematics},
volume = {58},
number = {4},
pages = {1094-1115},
year = {1998},
doi = {10.1137/S0036139996303877}  
}

@Article{Rand2003,
author={Rand, Richard
and Guennoun, Kamar
and Belhaq, Mohamed},
title={2:2:1 Resonance in the Quasiperiodic Mathieu Equation},
journal={Nonlinear Dynamics},
year={2003},
month={Mar},
day={01},
volume={31},
number={4},
pages={367-374},
issn={1573-269X},
doi={10.1023/A:1023216817293},
OPTurl={https://doi.org/10.1023/A:1023216817293}
}

@article{Kovacic2018,
    author = {Kovacic, Ivana and Rand, Richard and Mohamed Sah, Si},
    title = {Mathieu's Equation and Its Generalizations: Overview of Stability Charts and Their Features},
    journal = {Applied Mechanics Reviews},
    volume = {70},
    number = {2},
    pages = {020802},
    year = {2018},
    month = {02},
    issn = {0003-6900},
    doi = {10.1115/1.4039144},
    opturl = {https://doi.org/10.1115/1.4039144},
    opteprint = {https://asmedigitalcollection.asme.org/appliedmechanicsreviews/article-pdf/70/2/020802/6075492/amr_070_02_020802.pdf},
}

@Article{Zounes2002,
author={Zounes, Randolph S.
and Rand, Richard H.},
title={Global Behavior of a Nonlinear Quasiperiodic Mathieu Equation},
journal={Nonlinear Dynamics},
year={2002},
month={Jan},
day={01},
volume={27},
number={1},
pages={87-105},
issn={1573-269X},
doi={10.1023/A:1017931712099},
opturl={https://doi.org/10.1023/A:1017931712099}
}

@article{LAKRAD20041149,
title = {Effects of a low frequency parametric excitation},
journal = {Chaos, Solitons \& Fractals},
volume = {22},
number = {5},
pages = {1149-1164},
year = {2004},
issn = {0960-0779},
doi = {10.1016/j.chaos.2004.03.029},
opturl = {https://www.sciencedirect.com/science/article/pii/S0960077904001651},
author = {Faouzi Lakrad and Werner Schiehlen}
}

@article{LAKRAD20092426,
title = {Quasi-periodic solutions and periodic bursters in quasiperiodically driven oscillators},
journal = {Communications in Nonlinear Science and Numerical Simulation},
volume = {14},
number = {5},
pages = {2426-2433},
year = {2009},
issn = {1007-5704},
doi = {10.1016/j.cnsns.2008.09.007},
opturl = {https://www.sciencedirect.com/science/article/pii/S1007570408003079},
author = {Faouzi Lakrad and Mohamed Belhaq}
}

@Article{Lakrad2016,
author={Lakrad, F.
and Chtouki, A.
and Belhaq, M.},
title={Nonlinear vibrations of a shallow arch under a low frequency and a resonant harmonic excitations},
journal={Meccanica},
year={2016},
month={Nov},
day={01},
volume={51},
number={11},
pages={2577-2587},
issn={1572-9648},
doi={10.1007/s11012-016-0470-7},
opturl={https://doi.org/10.1007/s11012-016-0470-7}
}

@Article{Chtouki2020,
author={Chtouki, A.
and Lakrad, F.
and Belhaq, M.},
title={Quasi-periodic bursters and chaotic dynamics in a shallow arch subject to a fast--slow parametric excitation},
journal={Nonlinear Dynamics},
year={2020},
OPTmonth={Jan},
day={01},
volume={99},
number={1},
pages={283-298},
issn={1573-269X},
doi={10.1007/s11071-019-05082-7},
OPTurl={https://doi.org/10.1007/s11071-019-05082-7}
}

@article{Melnikov1963,
  title={On the stability of a center for time-periodic perturbations},
  author={Melnikov,  VK},
  journal={Transaction of the Moskow Mathematical Society},
  volume={12},
  pages={3--52},
  year={1963},
}

@article{Yagasaki1990,
    author = {Yagasaki, Kazuyuki and Sakata, Masaru and Kimura, Koji},
    title = {Dynamics of a Weakly Nonlinear System Subjected to Combined Parametric and External Excitation},
    journal = {Journal of Applied Mechanics},
    volume = {57},
    number = {1},
    pages = {209-217},
    year = {1990},
    month = {03},
    issn = {0021-8936},
    doi = {10.1115/1.2888306},
    opturl = {https://doi.org/10.1115/1.2888306},
    opteprint = {https://asmedigitalcollection.asme.org/appliedmechanics/article-pdf/57/1/209/5460861/209_1.pdf},
}

@book{Guckenheimer2013,
  title={Nonlinear Oscillations, Dynamical Systems, and \\Bifurcations of Vector Fields},
  author={Guckenheimer, J. and Holmes, P.},
  isbn={9781461211402},
  lccn={82019641},
  series={Applied Mathematical Sciences}, 
  year={2013},
  address   = {New York},
  publisher={Springer}
}

@article{Mathieu1868,
  author  = {Mathieu, Emile},
  title   = {M{\'e}moire sur le mouvement vibratoire d'une membrane
             de forme elliptique},
  journal = {Journal de Math{\'e}matiques Pures et Appliqu{\'e}es},
  volume  = {13},
  pages   = {137--203},
  year    = {1868}
}

@article{Floquet1883,
  author  = {Floquet, Gaston},
  title   = {Sur les {\'e}quations diff{\'e}rentielles lin{\'e}aires
             {\`a} coefficients p{\'e}riodiques},
  journal = {Annales scientifiques de l'{\'E}cole Normale Sup{\'e}rieure},
  volume  = {12},
  pages   = {47--88},
  year    = {1883}
}

@article{Hill1886,
  author  = {Hill, George William},
  title   = {On the part of the motion of the lunar perigee which is a
             function of the mean motions of the sun and moon},
  journal = {Acta Mathematica},
  volume  = {8},
  pages   = {1--36},
  year    = {1886}
}

@article{LAKRAD2005813,
title = {Bursters and quasi-periodic solutions of a self-excited quasi-periodic Mathieu oscillator},
journal = {Chaos, Solitons \& Fractals},
volume = {24},
number = {3},
pages = {813-824},
year = {2005},
issn = {0960-0779},
doi = {10.1016/j.chaos.2004.09.097},
opturl = {https://www.sciencedirect.com/science/article/pii/S0960077904005818},
author = {Faouzi Lakrad and Abderrahim Azouani and Nazha Abouhazim and Mohamed Belhaq}
}

@article{Kartal2026,
    author = {Kartal, Enise and Shoshani, Oriel and Botnaru, Elena and Mart\'in-P\'erez, Alberto and Manzaneque, Tomás and Alijani, Farbod},
    title = {Frequency Stability
of Graphene Nonlinear Parametric
Oscillator},
    journal = {Nano Letters},
    volume = {26},
    number = {23},
    pages = {7501-7508},
    year = {2026},
    OPTmonth = {06},
    issn = {1530-6984},
    doi = {10.1021/acs.nanolett.6c00581},
    url = {https://doi.org/10.1021/acs.nanolett.6c00581},
    OPTeprint = {https://pubs.acs.org/nalefd/article-pdf/26/23/7501/66256527/acs.nanolett.6c00581.pdf},
}

@Article{Howell1997,
author={Howell, K. C.
and Barden, B. T.
and Lo, M. W.},
title={Application of Dynamical Systems Theory to Trajectory Design for a Libration Point Mission},
journal={The Journal of the Astronautical Sciences},
year={1997},
month={Jun},
day={01},
volume={45},
number={2},
pages={161-178},
issn={2195-0571},
doi={10.1007/BF03546374},
OPTurl={https://doi.org/10.1007/BF03546374}
}

@article{Houri2019,
  author  = {Houri, S. and Ohta, R. and Asano, M. and Blanter, Y. M. and Yamaguchi, H.},
  title   = {Pulse-width modulated oscillations in a nonlinear resonator under two-tone driving as a means for {MEMS} sensor readout},
  journal = {Jpn. J. Appl. Phys.},
  volume  = {58},
  pages   = {SBBI05},
  year    = {2019}
}

@article{Houri2020,
  title = {Generic Rotating-Frame-Based Approach to Chaos Generation in Nonlinear Micro- and Nanoelectromechanical System Resonators},
  author = {Houri, Samer and Asano, Motoki and Yamaguchi, Hiroshi and Yoshimura, Natsue and Koike, Yasuharu and Minati, Ludovico},
  journal = {Phys. Rev. Lett.},
  volume = {125},
  issue = {17},
  pages = {174301},
  numpages = {6},
  year = {2020},
  OPTmonth = {Oct},
  publisher = {American Physical Society},
  doi = {10.1103/PhysRevLett.125.174301},
  OPTurl = {https://link.aps.org/doi/10.1103/PhysRevLett.125.174301}
}

@article{Houri2021,
  title = {Self-Sustained Libration Regime in Nonlinear Microelectromechanical Devices},
  author = {Houri, Samer and Asano, Motoki and Okamoto, Hajime and Yamaguchi, Hiroshi},
  journal = {Phys. Rev. Appl.},
  volume = {16},
  issue = {6},
  pages = {064015},
  numpages = {12},
  year = {2021},
  OPTmonth = {Dec},
  publisher = {American Physical Society},
  doi = {10.1103/PhysRevApplied.16.064015},
  OPTurl = {https://link.aps.org/doi/10.1103/PhysRevApplied.16.064015}
}

@article{Defoort2021,
	author = {Defoort, Martial and Rufer, Libor and Fesquet, Laurent and Basrour, Skandar},
	date = {2021/02/19}, 
	doi = {10.1038/s41378-021-00241-6},
	id = {Defoort2021},
	isbn = {2055-7434},
	journal = {Microsystems \& Nanoengineering},
	number = {1},
	pages = {17},
	title = {A dynamical approach to generate chaos in a micromechanical resonator},
	OPTurl = {https://doi.org/10.1038/s41378-021-00241-6},
	volume = {7},
	year = {2021} }

@book{kryloff1947introduction,
  title={Introduction to Non-linear Mechanics},
  author={Kryloff, N. and Bogoliubov, N.N. and Lefschetz, S.},
  isbn={9780691079851},
  lccn={lc43001901},
  series={Annals of mathematics studies},
  OPTurl={https://books.google.nl/books?id=XvLHko4Ru3QC},
  year={1947},
  publisher={Princeton University Press}
}

@article{Bachtold2022,
	author = {Bachtold, Adrian and Moser, Joel and Dykman, M. I.},
	title = {{Mesoscopic physics of nanomechanical systems}},
	journal = {Reviews of Modern Physics},
	volume = {94},
	number = {4},
	pages = {045005},
	year = {2022},
	month = dec,
	issn = {1539-0756},
	publisher = {American Physical Society},
	doi = {10.1103/RevModPhys.94.045005}
}

@article{Ochs2021,
  author  = {Ochs, J. S. and Seitner, M. and Dykman, M. I. and Weig, E. M.},
  title   = {Amplification and spectral evidence of squeezing in the response of a strongly driven nanoresonator to a probe field},
  journal = {Phys. Rev. A},
  volume  = {103},
  pages   = {013506},
  year    = {2021},
  doi     = {10.1103/PhysRevA.103.013506}
}

@article{Huber2020,
  title = {Spectral {{Evidence}} of {{Squeezing}} of a {{Weakly Damped Driven Nanomechanical Mode}}},
  author = {Huber, J. S. and Rastelli, G. and Seitner, M. J. and K{\"o}lbl, J. and Belzig, W. and Dykman, M. I. and Weig, E. M.},
  year = {2020},
  OPTmont = jun,
  journal = {Phys. Rev. X},
  volume = {10},
  number = {2},
  pages = {021066},
  publisher = {{American Physical Society}},
  doi = {10.1103/PhysRevX.10.021066}
}
 %TC:ignore

  %\section*{Acknowledgments}   %\noindent

 \section*{Data availability}   \noindent
 Data supporting the findings in this manuscript are available upon request. 

 \section*{Competing interests}   \noindent
 We declare we have no competing interests.

\end{document}